\documentclass[10pt,fleqn,a4paper]{article}

\usepackage{graphicx}
\usepackage{epstopdf}
\usepackage{amssymb,amsthm,amsmath,mathrsfs}
\usepackage{multicol,color}
\usepackage{float}

\usepackage{booktabs}

\newcommand{\df}{\sl}
\newcommand{\cf}{{cf.\ }}
\newcommand{\eg}{{{\it e.g.}\ }}
\newcommand{\ie}{{{\it i.e.}\ }}

 \newcommand{\sca}{\varrho}
 \newcommand{\dee}{\mathrm{d}}

\newcommand{\re}{\mathop{\mathrm{Re}}}
\newcommand{\im}{\mathop{\mathrm{Im}}}
\newcommand{\e}{{\rm e}}
\newcommand{\I}{{\rm i}}
\newcommand{\R}{{\mathbb{R}}}
\newcommand{\C}{{\mathbb{C}}}
\newcommand{\Z}{{\mathbb{Z}}}

\newcommand{\T}{{\mathbb{T}}}

\newcommand{\EM}{{\cal E \!\:\!\! M}}

\newcommand{\eps}{\varepsilon}
\newcommand{\Hav}{\overline{H}}
\newcommand{\Gav}{\overline{G}}
\newcommand{\cH}{{\cal H}}
\newcommand{\cI}{{\cal I}}
\newcommand{\cO}{{\cal O}}
\newcommand{\deltah}{{\Delta h}}

\newcommand{\opin}[2]{\; ] #1, #2 [ \;}
\newcommand{\minus}[1]{\mbox{$ - #1$}}

\newtheorem{Theorem}{Theorem}
\newtheorem{Lemma}[Theorem]{Lemma}
\newtheorem{Proposition}[Theorem]{Proposition}
\newtheorem{Corollary}[Theorem]{Corollary}
\newtheorem{Definition}[Theorem]{Definition}
\newtheorem{Remark}[Theorem]{Remark}

\newenvironment{theorem}{\begin{Theorem} \begin{sl}}{\end{sl}
                         \end{Theorem}}
\newenvironment{lemma}{\begin{Lemma} \begin{sl}}{\end{sl} \end{Lemma}}

\begin{document}

\title{The 1:$\pm 2$ resonance}
\author{R.H. Cushman$^{1,3}$, Holger R. Dullin$^{2}$, Heinz Han{\ss}mann$^{1}$, Sven Schmidt$^2$\\
$^{1}$ Mathematisch Instituut, Universiteit Utrecht, \\
 Postbus 80.010, 3508 TA Utrecht, The Netherlands \\
$^{2}$ Department of Mathematical Sciences,\\
 Loughborough University, LE11 3TU, UK \\
 {\small H.R.Dullin@lboro.ac.uk} \\
$^{3}$ Department of Mathematics and Statistics,\\
 University of Calgary, Canada \\
}
\date{October 1, 2007}

\maketitle

\begin{abstract}
\noindent
On the linear level elliptic equilibria of Hamiltonian systems are
mere superpositions of harmonic oscillators.
Non-linear terms can produce instability, if the ratio of frequencies
is rational and the Hamiltonian is indefinite.
In this paper we study the frequency ratio
%$1/2$
$\pm 1/2$ and its unfolding.
In particular we show that for the indefinite
case ($1$:$-2$) the frequency ratio map in a neighbourhood of
the origin has a critical point, i.e.\ the twist condition is violated for
one torus on every energy surface near the energy of the equilibrium.
In contrast, we show that the frequency map itself is non-degenerate
(i.e. the Kolmogorov non-degeneracy condition holds) for every
torus in a neighbourhood of the equilibrium point.
As a by product of our analysis of the fequency map we obtain
%by product we are able to obtain
another proof of fractional monodromy in the $1$:$-2$ resonance.
\end{abstract}

\section{Introduction}
\label{introduction}

The problem of resonant equilibria in Hamiltonian systems is quite intricate.
%While \eg periodic orbits form $1$--parameter families, the equilibria of a
% generic Hamiltonian system are isolated.
The set of resonant frequencies is countable and dense within all frequencies.
It is therefore generic for a Hamiltonian system to have all equilibria
non--resonant, but the set of all such Hamiltonians is not open.
% (while still dense, being an intersection of countable many open and dense sets).
An approach that allows to recover an open and dense set of Hamiltonians is
to make a distinction between resonances of high and low order.
The latter concern equilibria with local dynamics significantly different
from that of any {\df Birkhoff normal form}.
In two degrees of freedom this makes the $0$:$0$, $0$:$1$, $1$:$\pm 1$,
$1$:$\pm 2$ and $1$:$\pm 3$~resonances low order resonances.
For high order resonances the local dynamics is essentially governed by
a Birkhoff normal form, \hbox{\cf \cite{sanders78}.}

Thus it
%It thus
requires an external parameter~$\lambda$ to encounter a
resonant equilibrium in a Hamiltonian system.
The r\^ole of~$\lambda$ is to detune the frequencies.
In fact, the $0$:$0$~resonance --- an equilibrium with nilpotent
linearization --- is of co--dimension~$2$ and needs generically
one detuning parameter for each of the two frequencies,
\hbox{\cf \cite{cushman87}.}
The $1$:$1$~resonance is special as well~: already the linear
centraliser unfolding needs $3$~detuning parameters.
In~\cite{cotter86} a universal unfolding of the full (nonlinear) problem
with $7$~parameters is computed.
The $0$:$1$~and $1$:$-1$~resonances lead under parameter variation to
hyperbolic eigenvalues.
This triggers the normally elliptic centre--saddle (or Hamiltonian
saddle--node) bifurcation~\cite{broer93} and the Hamiltonian Hopf
bifurcation~\cite{meer85}, respectively.

In this paper we consider both the $1$:$2$~and the $1$:$-2$~resonance~;
we use the notation ${\sigma } = \pm 1$ in order to treat these cases
in a parallel fashion.
%using $\sigma = \pm 1$ we treat these where possible in parallel.
Moving the resonant equilibrium to both the origin $(q, p) = 0$ of the
phase space~$\R^4$ and the origin
of parameter space to $\lambda = 0$,
%$\lambda =0$ of the parameter space
the linear part of the Hamiltonian vanishes and the quadratic part at
the resonance reads
\[
   H_2(q, p) \;\; = \;\; \frac{\omega}{2} \left[ p_1^2 + q_1^2
   \, + \, 2 \sigma (p_2^2 + q_2^2) \right]  \;.
\]
The eigenvalues are $\pm  \I \omega, \pm 2  \I \omega$ and thus
in $1$:$\pm 2$~resonance.
We scale the frequency $\omega$ to $1$ by a linear change of time.
The linear centraliser unfolding
\[
   G_2(q, p; \lambda) \;\; = \;\; \frac12 ({p_1^2 + q_1^2})
   \; + \; \sigma (p_2^2 + q_2^2) \; + \; \frac{\lambda}{2}
%   \left[ p_2^2 + q_2^2 \, - \, 2 \sigma (p_1^2 + q_1^2) \right]
\left(  p_1^2 + q_1^2  \right)
\]
consists in adding a convenient detuning term.

Our interest lies in the full nonlinear dynamics defined by a
family $H(q,p; \lambda)$ of Hamiltonians with quadratic part~$G_2$. In
addition to the linear terms we also set the constant term equal to zero.
Thus we are
%;next to the linear part we also set the constant part zero and are
confronted with
\begin{equation}
\label{fullfamily}
   H(q, p; \lambda) \;\; = \;\; G_2(q, p; \lambda) \; + \;
   G_3(q, p ; \lambda) \; + \; G_4(q, p ; \lambda) \; + \; \ldots
\end{equation}
where $G_k$ is a homogeneous polynomial of order~$k$ in~$(q, p)$
that may have $\lambda$--dependent coefficients.
Close to the origin $G_k$ dominates $G_l$ for $k<l$ and we claim
that already the cubic polynomial
\[
   H_3(q, p; \lambda) \;\; = \;\; G_2(q, p; \lambda)
   \; + \; G_3(q, p ; \lambda)
\]
contains the desired information, provided that the average $\Gav_3$
of $G_3$ over the periodic flow of $X_{H_2}$ does not
vanish at $\lambda = 0$.
Indeed, our first step will be to derive a normal form
%\footnote{HRD: why is $\lambda = 0$ in the following formula?}
%
\begin{equation}
\label{normalform}
   \Hav_3(q, p; \lambda) \;\; = \;\; G_2(q, p; \lambda)
   \; + \; \Gav_3(q, p ; \lambda) .
\end{equation}
%by averaging $G_3$ along the periodic flow generated by~$H_2$.
Note that this would also be the first step to compute a Birkhoff
normal form, for which the terms of odd order vanish.
Because of the resonance
%we do have an un-removeable
we have an unremoveable odd order term.
%and
The result is
%sometimes
called resonant Birkhoff normal form.

In case $\Gav_3$ does vanish identically we may proceed to
compute the Birkhoff normal form up to order 4,
%\[
%   \overline{\Hav}^4_{\lambda} \;\; = \;\; H^2_{\lambda}(q, p)
%   \; + \; \overline{\Hav}_4(q, p ; \lambda)
% \]
and for non--degenerate quartic terms this special
$1$:$\pm 2$~resonance becomes a high order resonance.
The tricky bit is when $\Gav_3$ does vanish at $\lambda = 0$,
but not identically in~$\lambda$.
This case is of co--dimension~$2$ and one would then use a coefficient
$c$ in~$\Gav_3$ as second unfolding parameter.

By construction the normal form Hamiltonian~\eqref{normalform} is
invariant under the periodic flow generated by~$H_2$.
Our second step
%will consist
consists in the reduction of this $S^1$--symmetry.
The origin $(q, p) = 0$ is fixed whence $S^1$ does not act freely and
we have to resort to singular reduction \cite{CushmanBates1997}.
This reduction is performed by choosing a {\df Hilbert basis}
$\{ \pi_1, \pi_2, \pi_3, \pi_4 \}$, \ie generators of the ring
of $S^1$--invariant functions as (global) co--ordinates.
A certain drawback is that this basis is not free, whence we have to
work on the $3$--dimensional semi--algebraic subvariety of~$\R^4$ that
is defined by
\begin{equation}
\label{Intro:Equ1}
    \pi_1^2 \pi_2 \; = \; \pi_3^2 + \pi_4^2  \,,
    \quad \pi_1 \geq 0  \,, \; \pi_2 \geq 0  \;.
\end{equation}
The normal form $\Hav_3$ induces a Hamiltonian $\cH$ on this reduced phase space.
Being of one degree of freedom the corresponding reduced Hamiltonian
system is readily
%analysed, we
analysed. We do this both globally on the symplectic
leaves of the reduced phase space and locally in well--chosen co--ordinates.
See Elipe et al \cite{Elipe2001, Elipe2005} for a similar approach in the
setting of stability analysis.
%and locally in well--chosen co--ordinates
%that unite the $1$:$2$~and the $1$:$-2$~resonance in one phase portrait.
Due to the geometry of the reduced phase space, it is convenient to plot the energy surface in the space of the invariants (figures \ref{reduction:fig2} and \ref{reduction:fig3}). This technique was, in fact, already known to Alfriend \cite{Alfriend1973} and even earlier to Breakwell and Pringle \cite{JohnVBreakwellRalphPringle1966}.

Our third step concerns the dynamics in two degrees of freedom.
This consists of two parts.
First we reconstruct the dynamics defined by the normal form
$\Hav_3$ from the reduced Hamiltonian~$\cH$.
In the case $\sigma = +1$ of the $1$:$2$~resonance this makes $\R^4$
a {\df ramified torus bundle}, with regular fibres $\T^2$ reconstructed
from the periodic orbits of the reduced system.
In the case $\sigma = -1$ of the $1$:$-2$~resonance the reduced system always has unbounded
%motion and
motion. It may also have bounded orbits depending on the detuning.
We show by a direct computation
that $\R^4$ becomes
%similarly
a ramified fibre bundle, where the
regular fibres are $2$--tori and cylinders reconstructed from bounded
and unbounded orbits, respectively, in the pre--image of the regular values of the
energy--momentum mapping.
%, respectively.
In order to give a global overview of the dynamics we then go on and construct
in both cases $\sigma = \pm 1$ the sets of critical values of the
energy--momentum mappings.

The fourth step relates the reconstructed integrable
dynamics defined by~\eqref{normalform} to the ``original'' dynamics
defined by~\eqref{fullfamily}.
For $\sigma = +1$ this perturbation analysis is straightforward.
Things turn out to be more interesting for the $1$:$-2$~resonance.
Indeed, we show that the ramified fibre bundle defined
by~\eqref{normalform} displays fractional monodromy around the
singular fibre containing the origin $(q, p) = 0$ (in the case
$\sigma = +1$ the corresponding singular fibre consists of the
origin alone).
This phenomenon is treated in Section~\ref{period}.
The main technical issue is to define the notion of
frequency and frequency ratio (which shows monodromy)
for the non-compact cylinders.
% instead of on compact tori.
The approach taken in \cite{CES} is to add higher
order terms to the Hamiltonian that make the energy surface
compact. We follow a different approach along the lines
of \cite{SanVuNgocFF} where only the singular part of the dynamics
near the equilibrium point is considered. This singular part
captures the dynamics up to a smooth (but unknown) function
that
%captures
encode the influence of higher order terms.
However, the singular terms dominate the asymptotic behaviour of the frequency map
when sufficiently close to the equilibrium, and thus nothing
needs to be known about the higher order terms, except
their existence and smoothness.

The unfolding shows that
%there is a difference to
our situation is different from the unfolded $1$:$-1$ resonance.
When passing through a $1$:$-1$~resonance, (non--fractional) monodromy
is created in a supercritical Hamiltonian Hopf bifurcation and local
monodromy is turned into non-local \cite{WDR03} island  \cite{ES04} monodromy in the subcritical case.
%add reference to Efstathiou island monodromy.
The situation is different for the $1$:$-2$~resonance (where no
distinction super/subcritical exists).
Fractional monodromy \cite{NekhoroshevSadovskiiZhilinskii2006} is local at $\lambda = 0$ and turns into island monodromy for both positive and negative values of~$\lambda$.
Stated differently, the system has fractional monodromy before and after resonance.
%already has fractional monodromy, and when
When passing through a $1$:$-2$~resonance
%this
island monodromy momentarily contracts to local monodromy.

In the final section we come to
%the
our main new results on the 1:$-2$ resonance.
% \footnote{HRD: do we have anything about the frequency maps for the 1:$+2$ resonance?}
We show that the isoenergetic non-degeneracy (or twist) condition
is violated for a single torus on every energy near the critical energy.
This shows that the principle ``low order resonance leads to vanishing twist"
(already established for low order resonant fixed points of maps \cite{HRDMeiss,HRDIvanovSN})
is also valid for low order resonant equilibria.
%It also
This shows that
the principle ``monodromy causes vanishing twist", which has been
established for the Hamiltonian Hopf bifurcation \cite{HRDullinAVIvanov2005}
and for focus-focus equilibria in general \cite{HRDSanVuNgoc} also holds
in the more degenerate (and more difficult) situation of the 1:$-2$ resonance.
The second main new result is that the frequency map for tori
near the 1:$-2$ equilibrium is non-degenerate, i.e.\ the Kolmogorov condition
for the KAM theorem holds. Again, this is similar to the situation for
focus-focus points \cite{HRDSanVuNgoc}. The main difference
with the focus-focus equilibrium is that this
%is that the focus-focus
equilibrium was assumed to be non-degenerate (in the sense that the linearisation
of the commuting vector fields generate the Cartan sub-algebra); whereas in
the 1:$-2$ equilibrium it is always degenerate. Our analysis shows that this
implies that the period only has algebraic blowup, as opposed to the usual
logarithmic one.

Our results are valid for all completely integrable systems
that agree up to the cubic terms
with the normal form we analyse. The higher order terms do not change the asymptotic
properties of the frequency map when sufficiently close to the origin.
An arbitrary Hamiltonian system near the $1:-2$ resonance is generically non-integrable
due to higher order terms. Thus is does not posses a frequency map.
Using our results about the non-degeneracy of the frequency map, then KAM theory
can be invoked to show the persistence of tori whose frequencies are Diophantine.
Moreover,
%also
monodromy can be defined for the nearby non-integrable system \cite{HWBroer2007}.

\section{Normal form}
\label{normal}

The art of normalizing Hamiltonian Systems has been periodically (re)discovered and
forgotten and goes at least back to Birkhoff.
For treatments similar to ours see \eg~\cite{meer85, broer03a, HRDullinAVIvanov2005}.
The resonant quadratic part~$H_2$ generates the $S^1$--action
\[
\begin{array}{cccc}
   \varrho : & S^1 \times \C^2 & \longrightarrow & \C^2  \\
   & (t, z_1, z_2) & \mapsto & (z_1 \e^{\I t}, z_2 \e^{2 \sigma \I t} )
\end{array}
\]
where $z_k = (p_k + \I q_k)/\sqrt{2} \, , \, k = 1, 2$.
Introducing
\[
   z_2^{\sigma} \;\; = \;\; \left\{
   \begin{array}{c}  \bar{z}_2 \\ z_2  \end{array}
   \right. \;\; \mbox{if} \;\;
   \begin{array}{c}  \sigma = + 1 \\ \sigma = -1  \end{array}
\]
this action has the invariants
\[
   \pi_1 = z_1 \bar{z}_1 \,, \quad
   \pi_2 = z_2 \bar{z}_2 \,, \quad
   \pi_3 = \re( z_1^2 z_2^{\sigma}) \,, \quad
   \pi_4 = \im( z_1^2 z_2^{\sigma}) \;.
\]
These form a Hilbert basis of the ring $\left( C(\R^4) \right)^{\varrho}$
of $\varrho$--invariant functions, \ie given a $\varrho$--invariant function
$f : \R^4 \longrightarrow \R$ there exists $g : \R^4 \longrightarrow \R$
such that
\[
   f(q, p) \;\; \equiv \;\; g(\pi(q, p)) \;.
\]
Normalization of~$H(q,p;\lambda)$ with respect to this action gives a
Hamiltonian that is a function of the invariants only.
The only cubic polynomial that is a polynomial of the invariants is
the linear combination $a \pi_3 + b \pi_4$.
A simultaneous rotation in both the $(q_1, p_1)$--plane and the
$(q_2, p_2)$--plane turns this expression into $c \pi_3$ with $c^2= a^2 + b^2$.
Unless $c=0$ we
%may furthermore
assume that $c >0$.
%$c$ to be positive.
This yields the following result~\cite{duistermaat84a, kummer86}.

\begin{theorem}
Let the family $H(q,p;\lambda)$ of two--degree--of--freedom Hamiltonian systems
be given by~\eqref{fullfamily}, \ie with quadratic part unfolding the
1:$\pm 2$~resonance. Then there is a symplectic co--ordinate transformation~$\psi$
that turns $H(q,p;\lambda)$ into normal form
\[
   \left( H \circ \psi \right) (q, p; \lambda) \;\; = \;\;
   G_2(q, p; \lambda) \; + \; \Gav_3(q, p ; \lambda) \; + \;
   \mbox{\sl higher order terms}
\]
with
$\Gav_3(q, p ; \lambda) = c(\lambda) \left[
 (p_1^2 - q_1^2) p_2 + 2 \sigma p_1 q_1 q_2 \right]$.
\end{theorem}

\noindent
Introducing the scaling $(q, p, H) \mapsto (\eps q, \eps p, \eps^2 H)$
zooms into the origin and reveals the smallness of the higher order
terms as the normal form reads
\[
   \left( H \circ \psi \right) (q, p; \lambda) \;\; = \;\;
   G_2(q, p; \lambda) \; + \; \eps \Gav_3(q, p ; \lambda) \; + \;
   \cO \left( \eps^2 \right) \;.
\]
Restricting to a small interval $\opin{\minus{\lambda_0}}{\lambda_0}$
in parameter space defined by
\[
   c(\lambda) \;\; \equiv \;\; c(0) \; + \; \cO \left( \eps \right)
\]
we may furthermore put $c(\lambda) \equiv c(0) =: c$ in $\Gav_3$
whence the normal form~\eqref{normalform} turns out to be an
$\eps^2$--small perturbation of~\eqref{fullfamily}.
Note that
%for
if $c^{\prime}(0) \neq 0$,
%we have
then $\lambda_0 = \cO(\eps)$,
%~;we therefore
Therefore we may rescale $\lambda \mapsto \eps \lambda$ as well.
%and thus turn
This turns \eqref{normalform} into
\begin{eqnarray*}
   \Hav_3(q, p;\lambda ) & = & \frac12 ({p_1^2 + q_1^2})
   \; + \; \sigma (p_2^2 + q_2^2)  \\
    & & \!\!\!\! \mbox{} + \; \eps \left\{ \frac{\lambda}{2}
%   \left[ p_2^2 + q_2^2 \, - \, 2 \sigma (p_1^2 + q_1^2) \right] \: + \:
   \left( p_1^2 + q_1^2 \right) \: + \:
   c \left[ (p_1^2 - q_1^2) p_2 + 2 \sigma p_1 q_1 q_2 \right] \right\} .
\end{eqnarray*}
%and assume from now on
From now on we assume that the non--degeneracy condition $c > 0$ holds.
%to hold.
The additional scaling $(\lambda, \eps) \mapsto (\lambda c, c^{-1} \eps)$
would yield $c=1$.
We keep $c$ in the calculation so that
we can absorb additional constants when they appear.
Moreover, later on we want to (re)-introduce the scaling
in a slightly different way in the image of the energy-momentum map.

\section{Reduction}
\label{reduction}

% \newpage %???
\begin{figure}[p!]

\begin{minipage}[t]{\textwidth}

% \begin{figure}[H]
\begin{multicols}{2}

\begin{center}
\includegraphics[width=0.45\textwidth]{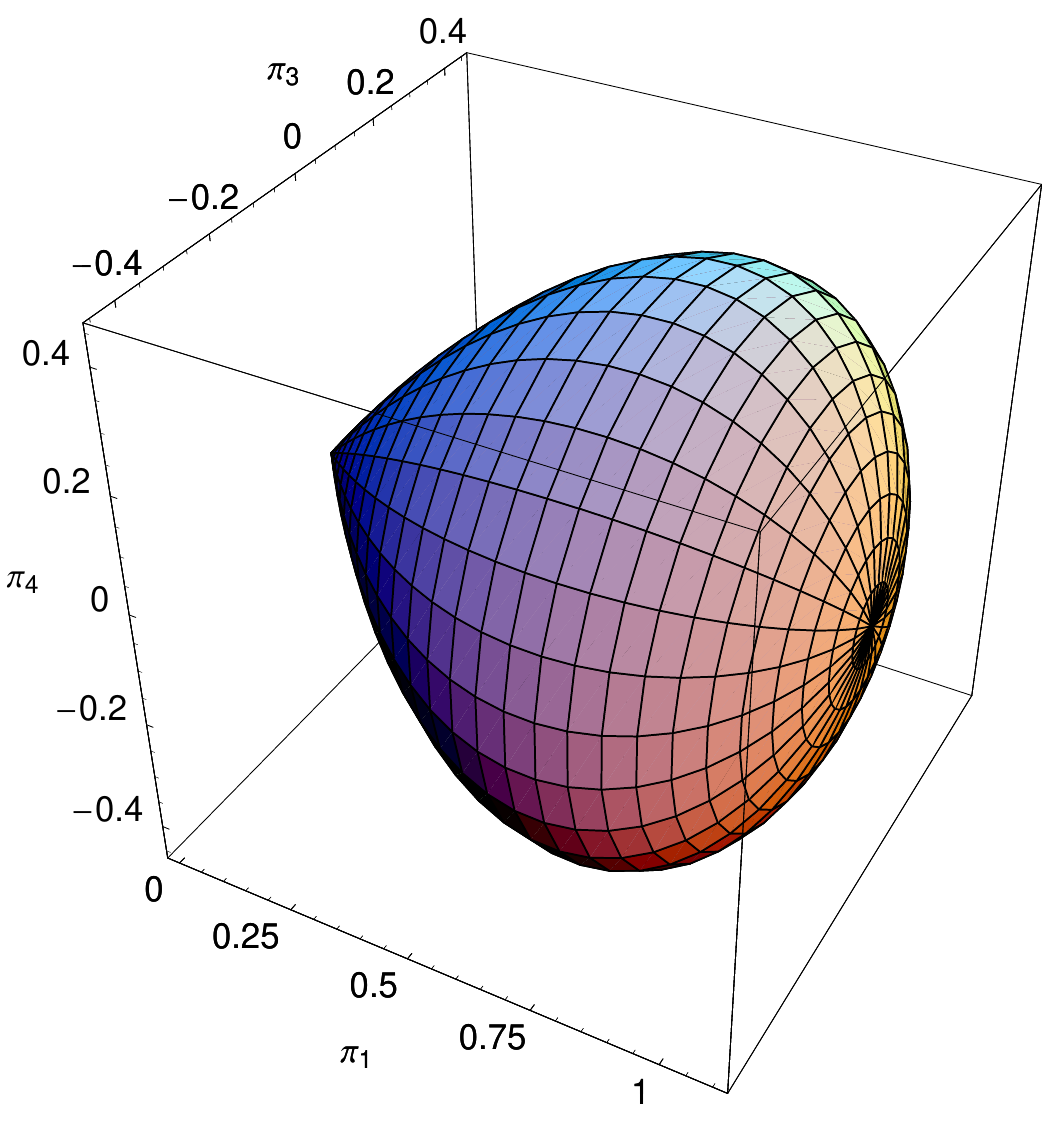}
\end{center}

\newpage

\begin{center}
\includegraphics[width=0.45\textwidth]{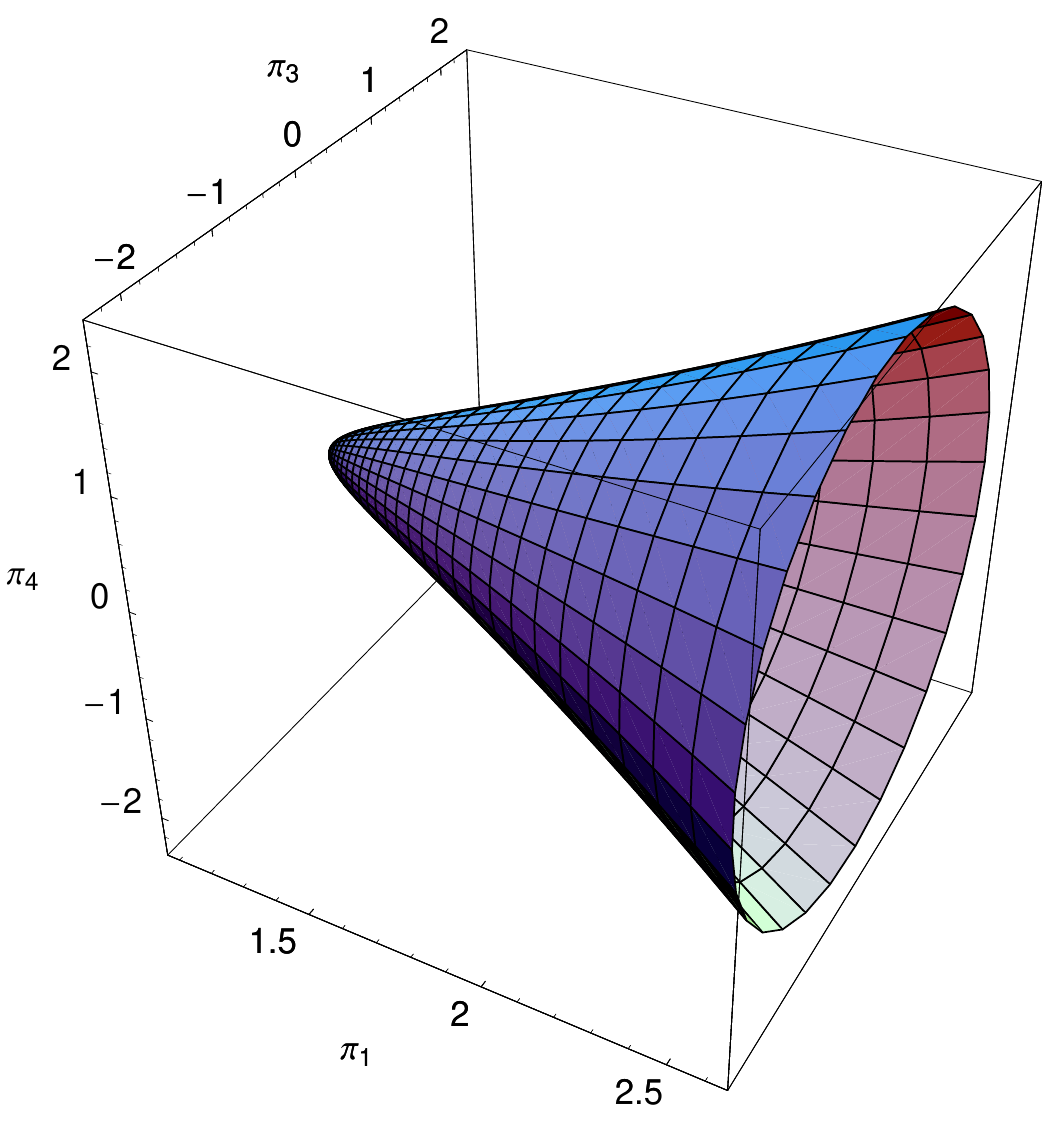}
\end{center}

\end{multicols}

\begin{multicols}{2}

\begin{center}
\includegraphics[width=0.45\textwidth]{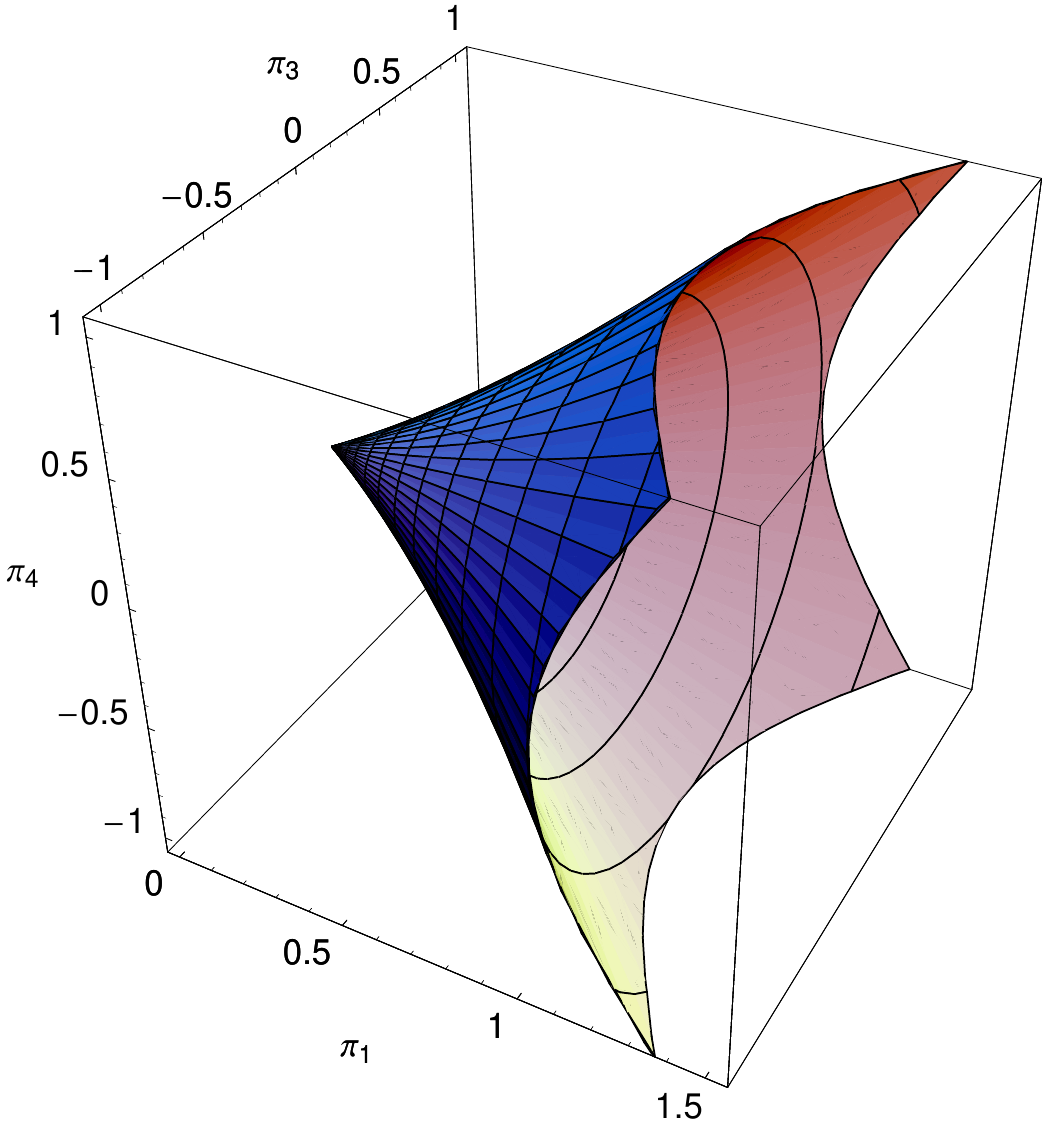}
\end{center}

\newpage

\begin{center}
\includegraphics[width=0.45\textwidth]{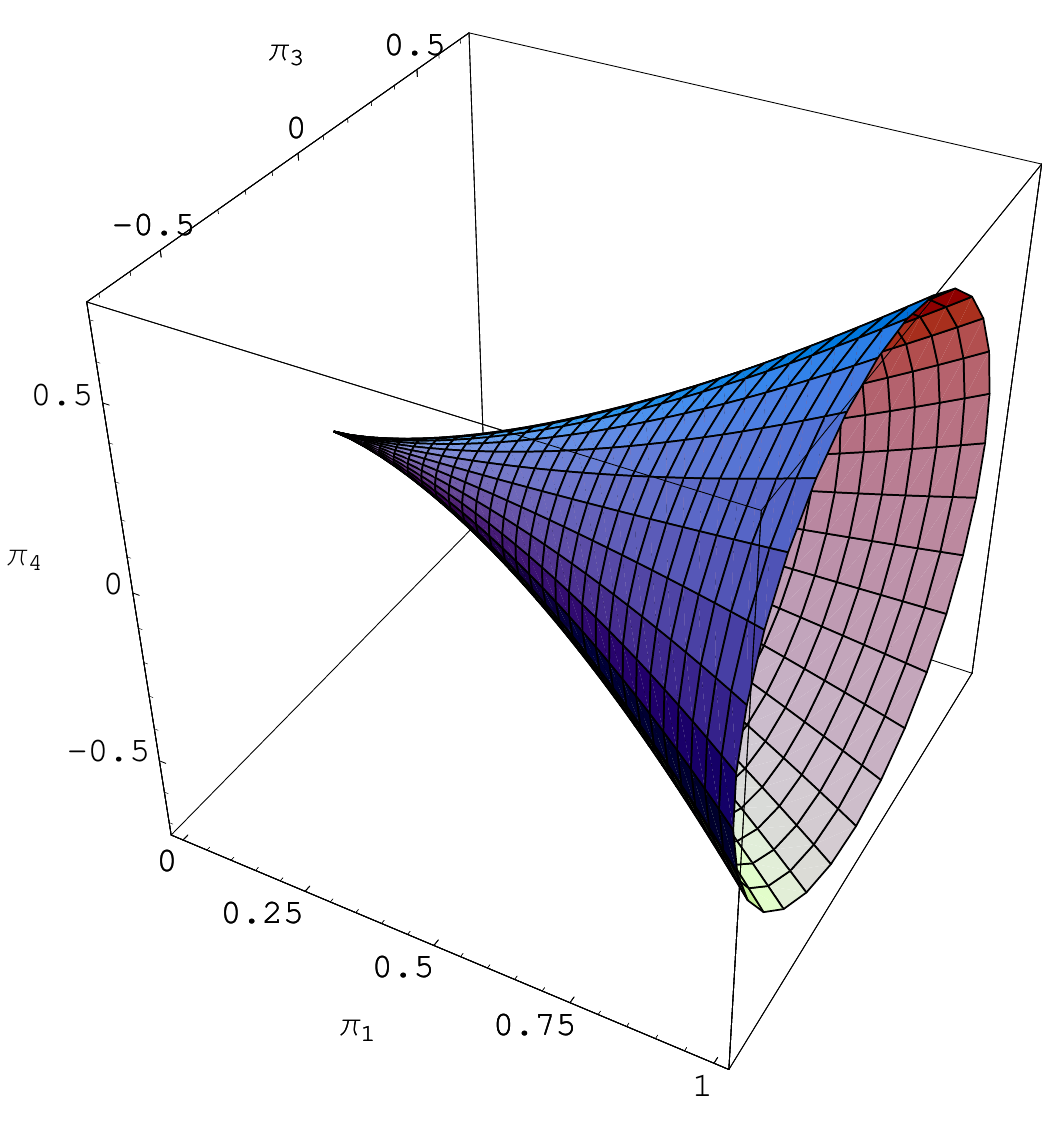}
\end{center}

\end{multicols}

\caption{\label{reduction:fig1}
\noindent
The reduced phase spaces that occur in the $1: \pm 2$ resonance.
Upper left: $\sigma=+1$ and $h_2 > 0$ (not shown: $P_{h_2}^{\sigma} = \emptyset$
for $\sigma = +1$ and $h_2 < 0$ and $P_0^+ = \{ 0 \}$);
upper right: $\sigma=-1$ and $h_2 > 0$, lower left: $\sigma=-1$ and $h_2 < 0$
and lower right: $\sigma=-1$ and $h_2 = 0$.
}

% \end{figure}

\end{minipage}

\end{figure}

% \newpage

The normalized truncated cubic Hamiltonian is Liouville integrable
with integral~$H_2$.
The flow generated by $H_2$ is periodic, and hence it is
the momentum of an $S^1$--action.
Singular reduction of this symmetry gives a system with one
degree of freedom.
Instead of working with $\pi_1, \pi_2, \pi_3, \pi_4$ it is
convenient to use a slightly different Hilbert basis of the ring
$\left( C(\R^4) \right)^{\varrho}$ of $\varrho$--invariant
functions that contains the generator of the $S^1$ action.
%We therefore eliminate
Therefore we drop $\pi_2$ in favour of
\[
\begin{array}{rcl}
   \eta & = &  \pi_1 \; + \; 2 \sigma \pi_2 \;\; = \;\;
   \frac{1}{2} (p_1^2 + q_1^2) \; + \; \sigma (p_2^2 + q_2^2)  \\
%   \pi_1 & = &  \pi_1 \;\; = \;\;
%   \frac{1}{2} (p_1^2 + q_1^2)  \\
%   \pi_3 & = & \pi_3 \;\; = \;\;
%   [(p_1^2 - q_1^2) p_2 \; + \;  2 \sigma p_1 q_1 q_2 ]/2\sqrt{2} \\
%   \pi_4 & = &  \pi_4 \;\; = \;\;
%   [2p_1 q_1 p_2 \; - \; \sigma (p_1^2 - q_1^2) q_2 ] /2\sqrt{2} .
\end{array}
\]
Whence the relation between the invariants reads $R_{\eta}(\pi_1, \pi_3, \pi_4) = 0$ with
\[
   R_{\eta}(\pi_1, \pi_3, \pi_4) \;\; = \;\;
%     (\eta-2\sigma\pi_1)^2 \pi_1   \; - \;  \pi_3^2 + \pi_4^2
    \frac\sigma2  \pi_1^2 (\eta- \pi_1)    \; - \;  (\pi_3^2 + \pi_4^2);
\]
while the inequalities $\pi_1 \geq 0, \pi_2 \geq 0$ turn into an
interval $\cI_{\eta}$ of admissible values for~$\pi_1$.
In the case $\sigma = +1$ of the 1:2~resonance this interval
is bounded and given by
\[
   \cI_{\eta}^+ \;\; = \;\;
   % \left[ - 2 \eta,   \textstyle{\frac{1}{2}} \eta \right]
   \left[ 0,   \eta \right] ;
\]
%while
whereas for $\sigma = -1$, the invariant $\eta$ may assume negative
values as well and
\[
   \cI_{\eta}^- \;\; = \;\;
%   \left[ \max \{ - \textstyle{\frac{1}{2}} \eta, 2 \eta \}   \, , \, \infty \right[
   \left[ \max \{ 0, \eta \}   \, , \, \infty \right[
\]
is unbounded.
The Poisson bracket relations between the invariants are easily computed
and given in table~\ref{table1}, where
\[
   f(\pi_1) \;\; = \;\;
   \frac\sigma2( 2\eta \pi_1 - 3 \pi_1^2)
   \;\; = \;\; \frac{\partial R_{\eta}}{\partial \pi_1} \;.
\]
Because $\eta$ coincides with $H_2$ it was to be expected that this
invariant is a Casimir of the reduced Poisson structure.

%\begin{table}[b!]
\begin{table}[hb!]
\begin{center}
\caption{ {\small The reduced Poisson bracket. \label{table1}}}
\begin{tabular}{c|cccccc}
   $\{ \uparrow , \leftarrow \}$
   & $\pi_1$ & $\pi_3$ & $\pi_4$ & $\eta$ \\ \hline
   $\pi_1$ & $0$ & $ 2 \pi_4$ & $ -2 \pi_3$ & $0$  \\
   $\pi_3$ & $-2\pi_4$ & $0$ & $f(\pi_1)$ & $0$  \\
   $\pi_4$ & $2 \pi_3$ & $-f(\pi_1)$ & $0$ & $0$  \\
   $\eta$ & $0$ & $0$ & $0$ & $0$
\end{tabular}
\end{center}
\end{table}

Fixing the value $h_2$ of~$\eta$ the one--degree--of--freedom
dynamics takes place in the surface of revolution (see figure~\ref{reduction:fig1})
\begin{displaymath}
   P_{h_2}^{\sigma} \;\; = \;\; \left\{ \; (\pi_1,\pi_3,\pi_4) \in \R^3 \; \mid \;
   R_{h_2}(\pi_1,\pi_3,\pi_4) = 0 \, , \, \pi_1 \in \cI_{h_2}^{\sigma} \; \right\}
\end{displaymath}
with Poisson structure
\begin{displaymath}
   \{ f, g \} \;\; = \;\;
   ( \nabla f \times \nabla g \mid \nabla R_{\ell} ) \;.
\end{displaymath}
Note that $P_{h_2}^+$ is compact and $P_{h_2}^-$ is unbounded.
The conical singular point at $\pi_1 =  0$ (unless $\sigma = -1$ and $h_2 > 0$)
comes from the isotropy subgroup $\Z_2$ of the $S^1$-action $\rho$ at $(0,z_2)$.
When $\sigma = -1$, such points satisfy $h_2 \leq 0$.

\begin{figure}
\centerline {\includegraphics[width=0.45\textwidth]{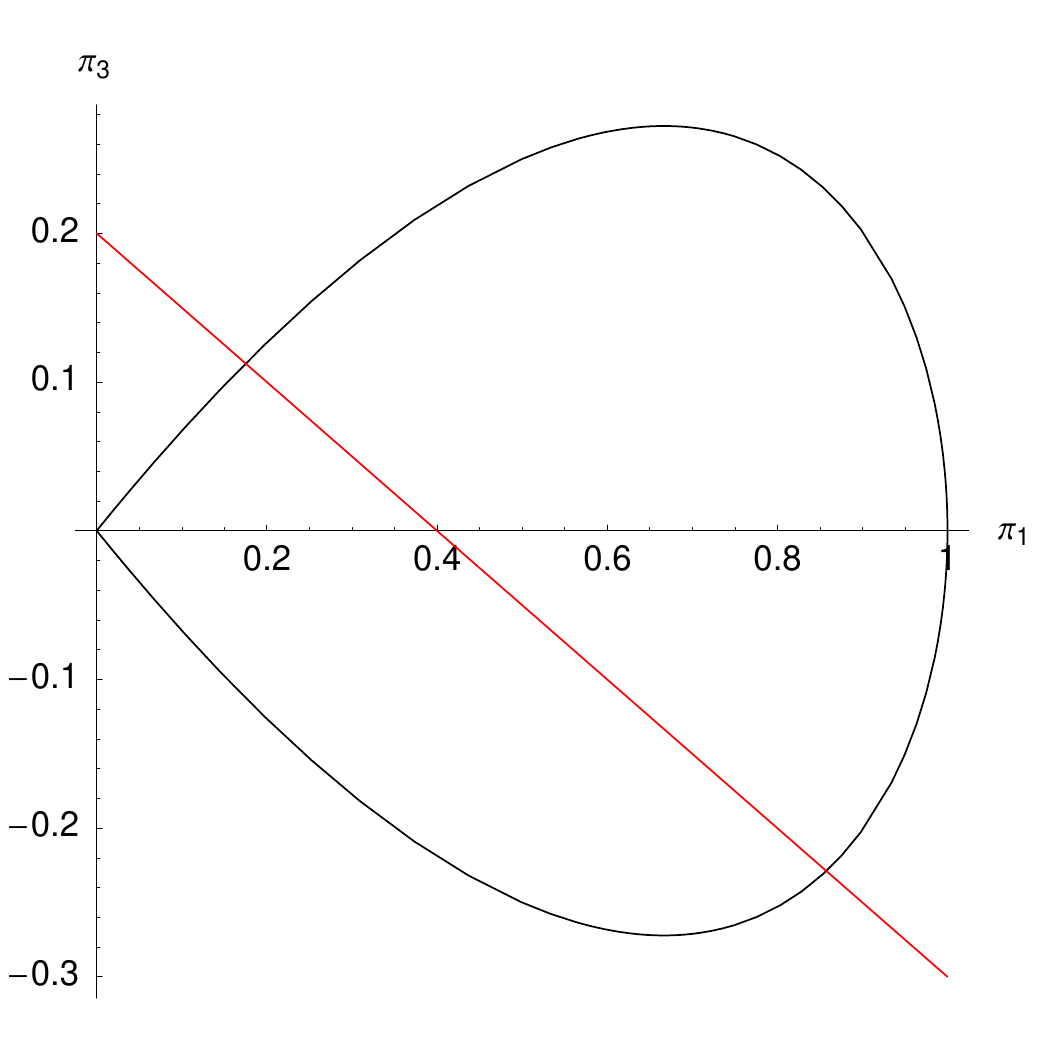}}
\caption{\label{reduction:fig2}
{\small Section $\pi_4 = 0$ and $\pi_1 + 2 \pi_2 = h_2$ of the reduced phase space for $\sigma = +1$, $h_2 = 1$, together with a line of constant Hamiltonian $h=h_2 + \lambda \pi_1 + c \pi_3 = 1.2$, $\lambda = 1/2$, $c=1$. }
%the notation $\deltah$ is not defined
}
\end{figure}

\begin{figure}
\centerline {\includegraphics[width=0.65\textwidth]{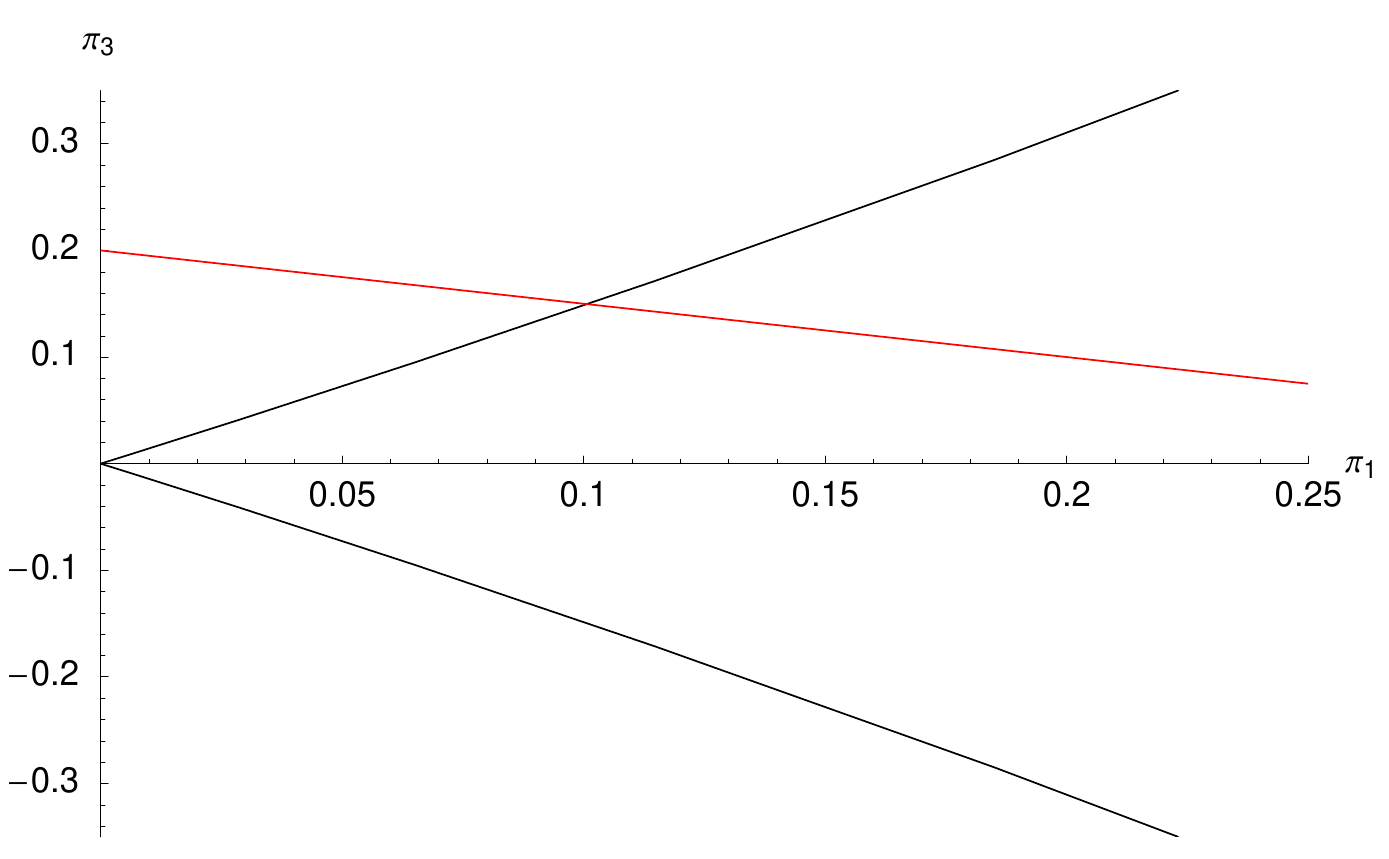}}
\centerline {\includegraphics[width=0.65\textwidth]{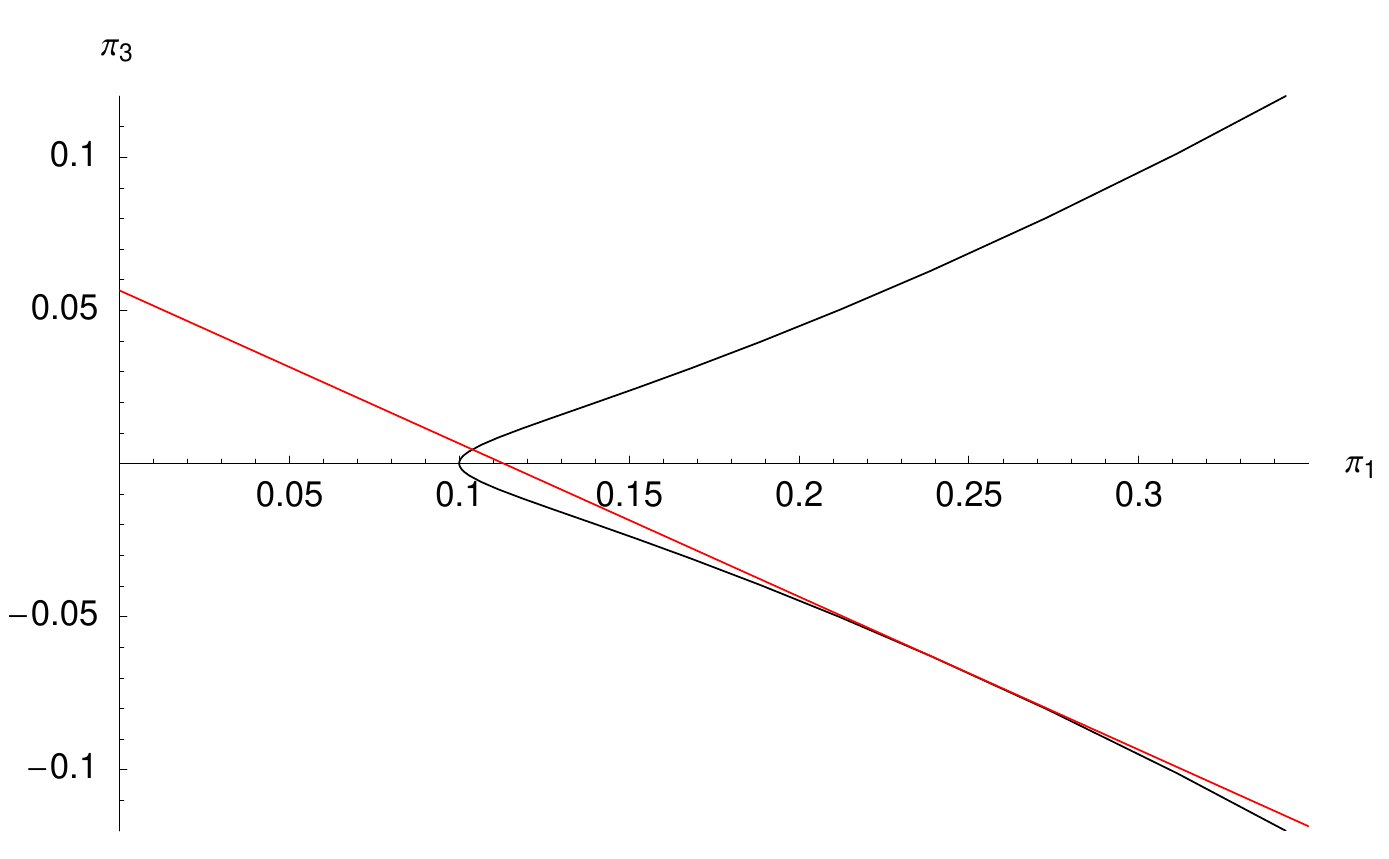}}
\caption{\label{reduction:fig3}
{\small Section $\pi_4 = 0$ and $\pi_1 - 2 \pi_2 = h_2$ of the non--compact reduced phase space for $\sigma = -1$.
For $h_2 = -1$ (top), there is a singular point at the origin.
The line of constant Hamiltonian is determined by $\deltah = 0.2$, $\lambda = 1/2$, $c = 1$.
For $h_2 = 0.1$ (bottom), the reduced phase space is smooth.
The line of constant Hamiltonian is given by $\deltah=0.0565$, $\lambda = 1/2$, $c=1$.}
}
\end{figure}

The reduced Hamiltonian is obtained by expressing~$\Hav_3$ in the
basic invariants and reads
\begin{equation}
\label{Reduction:Equ1}
  %  \hat{\Hav}
  \cH_3(\pi_1,\pi_3,\pi_4; \lambda) \;\; = \;\; \eta \; + \;
   \eps \left\{ \lambda \pi_1 + c \pi_3 \right\} \;,
\end{equation}
where the factor $2 \sqrt{2}$ has been absorbed in the factor~$c$.
Subtracting the constant value $h_2$ of $\eta$ and rescaling time by~$\eps$
yields
\[
   \cH(\pi_1,\pi_3,\pi_4; \lambda) \;\; = \;\; \lambda \pi_1 + c \pi_3 \;.
\]
It is important to keep in mind that subtracting $h_2$ is only a valid operation
in the reduced system, but not in the full system. Altering the Hamiltonian by
a constant in general makes no difference to the dynamics.
%, however, even though $H_2$
Eventhough $H_2$ is constant along trajectories
%in
of the full averaged system, it does
%full system as well, it does
contribute leading order terms to the equations of motion, and thus influences
the frequencies and rotation number.
In the reduced system the function $\eta$ (with value $h_2$) is a Casimir of the
reduced bracket. Hence removing it does not change the equations of motion.
So even though in the unreduced system we should not remove
$H_2$, we will see that the fractional monodromy is created by the non-linear
terms $\Gav_3$, and the monodromy is not changed by removing $H_2$.

The reduced integral curves are given by the intersections
\[
   P_{h_2}^{\sigma} \cap \cH^{-1}(\deltah) \, ,
\]
where $\deltah = h - h_2$ is the difference between the value~$h$ of the
energy and the value $h_2$ of~$\eta$, i.e.\ the value of the cubic terms
in the Hamiltonian.
%Next
In addition to the singular points, we have regular equilibria where
$\cH^{-1}(\deltah)$ is tangent to~$P_{h_2}^{\sigma}$.
Because the latter is a surface of revolution and the former is
a plane parallel to the $\pi_4$--axis this requires $\pi_4 = 0$.
Hence, we merely have to compute the points where the straight line
\[
   \pi_3 \;\; = \;\; \frac{\deltah}{c} \; - \; \frac{\lambda}{c} \pi_1
\]
is tangent to the cubic curve when ${\pi }_1 \in {\mathcal{I}}^{\sigma }_{h_2}$ and
\begin{equation}
\label{reduction:equ1}
   \pi_3^2 \;\; = \;\;
   \frac\sigma2  \pi_1^2 (\eta- \pi_1) ,
\end{equation}
%at a point $\pi_1 \in \cI_{h_2}^{\sigma}$,
see figures~\ref{reduction:fig2}
and~\ref{reduction:fig3}.
%In case this coinciding derivative occurs
When this happens at the singular point
${\pi}_1 = 0$, a Hamiltonian flip bifurcation occurs as $h_2$ is varied, cf.~\cite{h1h07}.
In the full system this corresponds to a supercritical period--doubling bifurcation.
This explains the phase portraits when $\sigma = +1$.
For $\sigma = -1$ the subcritical Hamiltonian flip bifurcation at the singular equilibrium
for $h_2 <0$ is accompanied by a centre--saddle
bifurcation (of regular equilibria) for $h_2 > 0$, both emanating from the
non--conical singularity $\pi_1 = 0$ at $\lambda = 0$.
This is best illustrated in terms of the critical values in the
image of the energy-momentum map, see below.

\subsection{Rational Parametrisation}

An interesting feature of the 1:$\pm 2$~resonances is that the two
semi--algebraic varieties $P_{h_2}^+$ and~$P_{-h_2}^-$ together form
the cubic~$R_{h_2}^{-1}(0)$.
Only when $h_2 < 0$ the reduced phase space~$P_{h_2}^+$ is empty and the
cubic~$R_{h_2}^{-1}(0)$ has an additional isolated point at the origin.
Since this cubic always has a singular point there exists a rational
parametrisation given by
\[
   (\pi_1, \pi_3)(s) \;\; = \;\;
%   ( h_2 \, - \, \sigma s^2, s (h_2 - \sigma s^2) )
  ( h_2 \, - \, 2 \sigma s^2, s \pi_1 )
\]
where $s = \pi_3/\pi_1$ is the slope of the line through the origin that
intersects the cubic at $(\pi_1, \pi_3)$.
Similarly there is an explicit polynomial parametrisation using the inhomogeneous
co--ordinates $s_1 = \pi_3/\pi_1$ and $s_2 = \pi_4/\pi_1$ such that
\[
  (\pi_1, \pi_2, \pi_3, \pi_4)(s_1, s_2) \;\; = \;\;
  ( h_2 \, - \, 2 \sigma (s_1^2 + s_2^2), s_1^2 + s_2^2, s_1 \pi_1, s_2 \pi_1 ) \;.
\]
This stereographic projection from the singular point onto the plane $\pi_1 = 1$
gives co--ordinates on the reduced phase space.
Here $(\pi_1,\pi_3, \pi_4)$ are like homogeneous co--ordinates,
while $s_1, s_2$ are inhomogeneous co--ordinates on the plane $\pi_1 = 1$.
A similar construction can be done with the plane $\pi_2=1$.

\begin{theorem}
The inhomogeneous co--ordinates
$(s_1, s_2) = (\pi_3/\pi_1, \pi_4/\pi_1)$ are symplectic
co--ordinates (up to a multiplier) on the singular reduced phase space:
\[
   \{ s_1, s_2 \} \; = \; \frac{ \sigma }{2} \;.
\]
The reduced Hamiltonian
$\cH(\pi; \lambda) = \lambda \pi_1 + c \pi_3$
in these co--ordinates is
\[
   \cH(s; \lambda) \;\; = \;\; (\lambda + c s_1)
   ( h_2 \, - \, 2 \sigma (s_1^2 + s_2^2) ) \;.
\]
\end{theorem}

\begin{proof} By direct calculation.
\end{proof}

\begin{figure}
\centerline{\includegraphics{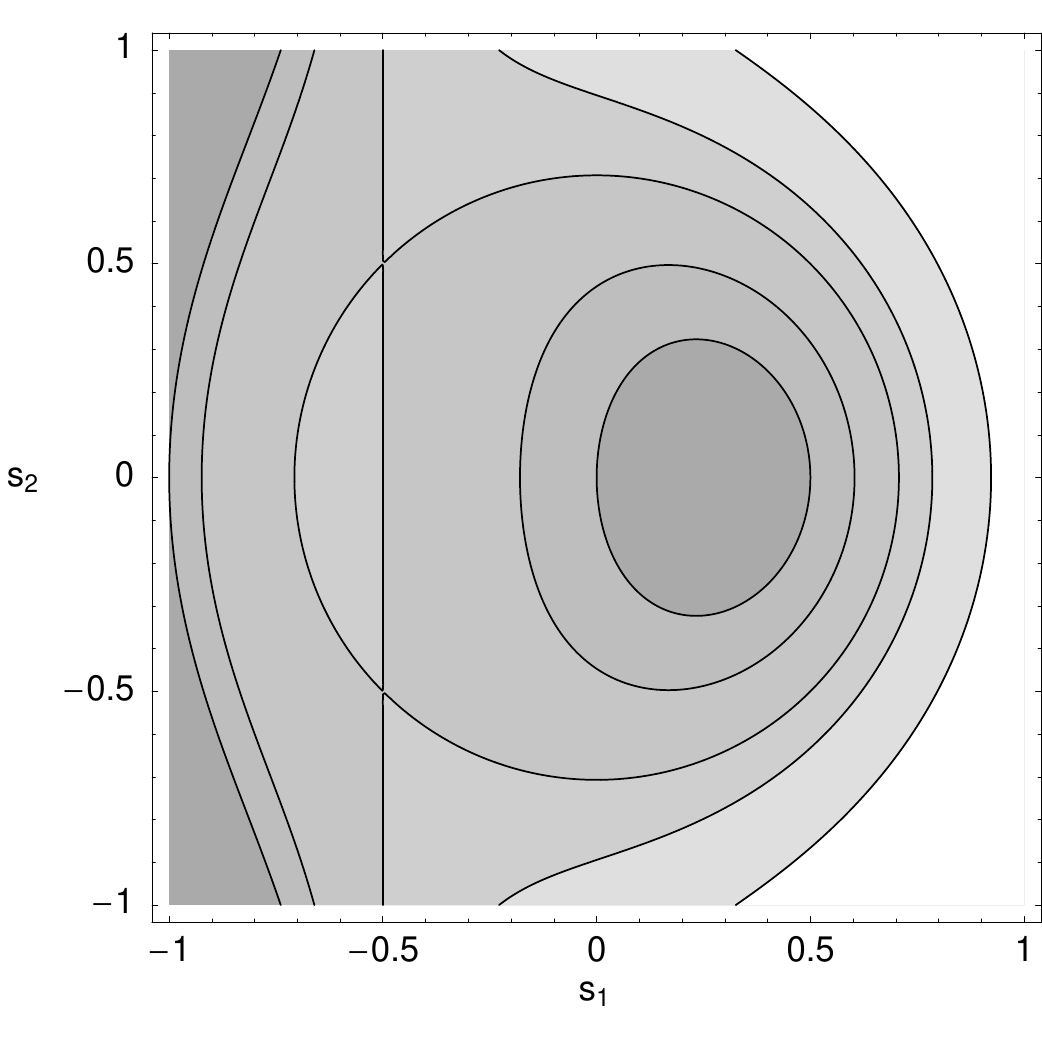}}
\caption{
{\small Phase portrait of $\cH(s;\lambda)$ for $\sigma = \pm 1$, $h_2 = \pm 1$, $\lambda = 1/2$, $c=1$. }
}
\label{cPP}
\end{figure}

\noindent
The level lines of $\cH(s;\lambda)$ define the planar phase portrait,
see figure~\ref{cPP} for an example.
The singular point of the reduced phase space (in $\pi_i$) is blown up to
the invariant circle $h_2 = \sigma (s_1^2 + s_2^2)$ in the phase portrait.
If the line $s_1 = -\lambda/c$ intersects the circle it contains two equilibrium points.
For $\sigma = 1$ the relevant part of the phase portrait is the closed disk
inside this circle, while for $\sigma = -1$ and $h_2 < 0$ it is the closure of
the complement. Recall that the two reduced phase spaces for $\sigma h_2 > 0$ are just
different parts of the same singular cubic curve, and in the same way
% all of the plane
the union of the two phase portraits gives all of the plane $\R^2 = \{ s_1, s_2 \}$.
% Changing the sign of $h_2$ and the sign of $\sigma$ does not change the contours.
%\\
\par
When there is no singular point in the reduced phase space, $\sigma h_2 < 0$,
the phase portrait is all of $\R^2 = \{ s_1, s_2 \}$.
The circle seen in the case $\sigma h_2 > 0$ now has negative radius.

The other prominent feature of the phase portrait is the invariant line $s_1 = -\lambda/c$.
It is the image of the energy surface $\deltah = 0$, which is a plane through the
origin $0 = \lambda \pi_1  + c \pi_3$ intersected with the reduced phase space.
When $\sigma h_2 > 0$ this line is the pre--image of the critical value.
If this line intersects the critical circle, then the equilibrium is unstable
and the line is the separatrix. If the line and the circle do not intersect,
or if there is no circle ($\sigma h_2 < 0$), then the line does not contain critical points, see Haller and Wiggins \cite{SWiggins1996}.

%Finally notice that when $\lambda = 0$ and $\sigma h_2 < 0$ there is a symmetry
%$s_1 \to -s_1, t \to -t$ in the Hamiltonian. This will become important for the computation
%of vanishing twist, see sec.~\ref{twist}. \footnote{HRD: really?}

\section{Dynamics in two degrees of freedom}
\label{dynamics}

To reconstruct the dynamics of~$\Hav_3$, we have to attach a
$1$--torus~$S^1$ to every regular point of~$P_{h_2}^{\sigma}$.
In this way the periodic orbits in~$P_{h_2}^{\sigma}$ give rise to
invariant $2$--tori and the unbounded reduced motions that exist
for $\sigma = -1$ give rise to invariant cylinders; while
%the
regular equilibria lead to periodic orbits in~$\R^4$.
For the singular equilibria the isotropy group becomes important.
%Therefore
The singular equilibria
$(\pi_1,\pi_2,\pi_3,\pi_4) = (0, h_2 / (2 \sigma), 0,0) \in P_{h_2}^{\sigma}$
with $h_2 \neq 0$ lead to periodic orbits as well, albeit with
half the period.
From the singular equilibria
%$\pi = 0$
${\pi }_1 =0 $, we reconstruct
the initial equilibrium at the origin of~$\R^4$.

The bifurcations of equilibria of the reduced system translate to
bifurcations of periodic orbits in two degrees of freedom.
The centre--saddle bifurcation turns straightforwardly into a
periodic centre--saddle bifurcation.
Where the singular equilibrium undergoes a Hamiltonian flip bifurcation
we obtain a period--doubling bifurcation, supercritical (new orbit is elliptic)
for $\sigma = +1$ and subcritical (new orbit is hyperbolic) for $\sigma = -1$.
Note that in the latter case the (hyperbolic) periodic orbit with twice
the period bifurcates off from the $\Z_2$--isotropic one as $h_2$ increases
through negative values (\ie $|h_2|$ {\em de}creases), while the elliptic
periodic orbit with twice the period comes into existence as $h_2$
increases through positive values.

In the latter case $\sigma = +1$ of the 1:2~resonance the periodic
orbits are elliptic and the regular fibres
of the energy--momentum mapping $\EM = (H_2, \Hav_3)$ are
compact.
%whence
Whence the foliation into $2$--tori is locally trivial and the
pre--image $\EM^{-1}(\bigcup \rho) \subseteq \R^4$ of regular values
$\rho$ of~$\EM$ is a $2$--torus bundle.
In the hyperbolic case the periodic orbits
%, in the hyperbolic case
together with their (un)stable manifolds, determine the structure of this bundle.
Together they turn the phase space~$\R^4$ into a ramified $2$--torus
bundle. A similar result for the case $\sigma = -1$ does not follow from
general theorems. We therefore show by direct computation that the
$1$:$-2$~resonance turns the phase space into a ramified
cylinder bundle.
%, and with
With detuning there are some compact tori
%around
as well.

\subsection{Local triviality}
\label{local}

Because $h = \deltah + h_2$ is linear in $\pi_3$ we can eliminate~$\pi_3$ from the
relation $R_{h_2} = 0$.
To this end we define the polynomial
\begin{equation} \label{Q}
    Q(z) = 2 \sigma c^2 z^2(h_2 -z) - 4 (\deltah -\lambda z)^2  \; ,
\end{equation}
whence $\pi_4^2 = Q(\pi_1)/(4c^2)$.
This is an elliptic curve, and can be parametrised by elliptic
functions.
Thus assume we have $( \pi_1(s; f ), \pi_4(s; f) )$ where $s \in \R$ is
the parameter
and $f = (h_2, \deltah)$ denotes the values of the energy--momentum mapping.
By the foregoing analysis we know that the curve
is not compact, but smooth when $f$ is a regular value.
By back-substiution we find
\begin{alignat*}{2}
   \pi_2(s, f) & = \frac{\sigma}{2} (h_2 - \pi_1(s; f)) \\
   \pi_3(s; f) & = \frac{1}{c}(h - h_2 - \lambda \pi_1(s; f)) \;.
\end{alignat*}
%Now
Next we need to find a point $(q,p)$ in phase space, so that the reduction mapping
$(q, p) \mapsto (\pi_1, \pi_2, \pi_3, \pi_4)$ sends it to
given values of the invariants. If such a mapping exists, we have a
%global
section of the bundle defined by the energy momentum mapping
in a neighbourhood of a regular value. This mapping is
\[
   (p_1, q_1, p_2, q_2) \;\; = \;\; \left(
   \sqrt{\pi_1}, 0, \frac{\pi_3}{\pi_1}, -\sigma \frac{\pi_4}{\pi_1}
   \right)  ,
\]
%This mapping
which is well defined because for regular values of $f$ the
invariant $\pi_1$ is nonzero (and positive in any case).
Finally, since the $S^1$
%action
acts on $(p,q)$ by rotation,
% and the
we have an explicit
parametrisation of the cylinders in phase space
%is
given by
% \footnote{HRD: check whether this $\phi$ is $\Theta$ used later on}
%
\begin{eqnarray*}
   & & (p_1, q_1, p_2, q_2) (s, \phi ; f) \;\; = \;\;  \\
   & & \qquad \left(
   \sqrt{\pi_1} \cos \phi, \sqrt{\pi_1} \sin \phi,
   (\pi_3\cos2\phi + \pi_4 \sin 2\phi)/\pi_1,
   \sigma (\pi_3\sin 2\phi - \pi_4 \cos2 \phi )/\pi_1
   \right) .
\end{eqnarray*}
The arguments $(s;f)$ for all $\pi_i$ have been suppressed.
This is the local trivialisation of the bundle defined by
the regular values of the energy momentum map.
%, such that the composition with the bundle projection (i.e.\ the energy--momentum
%mapping) gives back the base point~$f$.

\subsection{Energy--momentum mapping}
\label{enermom}

The following results on the structure of the bifurcation diagram for the
$1:\pm 2$ resonance were already described by Henrard \cite{henrard1970}. However, his discussion of $\sigma=-1$ was incomplete. Also see Schmidt and Sweet \cite{schmidtd1973}, Henrard \cite{henrard1973} and Sweet \cite{sweet1973} for a more general setting.

\begin{theorem}
The critical values of the energy--momentum mapping are contained in the set
given by the line $h = h_2$ and the cubic curve
\[
   [h_2, h - h_2] \;\; = \;\;
   \frac{\sigma}{2}( t - 2 \lambda) [ 3t - 2\lambda, t^2] / c^2 \;.
\]
For $t = 0$ the curve and the line $h = h_2$ are tangent at $[h_2, h-h_2] = \sigma[2 \lambda^2, 0]/c^2$ .\\
For $t = 4\lambda/3$ the curve has a cusp at $[h_2, h-h_2] = -2\sigma[9 \lambda^2, 8\lambda^3]/27/c^2$. \\
For $t = 2 \lambda$ the curve intersects the line $h=h_2$ transversely in the origin.
\end{theorem}
\begin{proof}
Eliminating $\pi_3$ in relation \eqref{reduction:equ1} by using the Hamiltonian gives the polynomial
$Q$ defined in the previous section in (\ref{Q}).
Tangencies of the reduced Hamiltonian with the reduced phase space
are given by the double roots of the cubic polynomial $Q$.
Equating coefficients in $ - \sigma Q(z)/(2 c^2) - (z - r)^2(z + \sigma t^2/ (2 c^2) ) = 0$ gives the parametrisation of the cubic discriminant surface where
$\pi_1 = r = \sigma t(t-2\lambda)/c^2$. A second solution branch with $r = 0$ and $h = h_2$
for which $h_2 = \sigma (4\lambda^2 - t^2)/2/c^2$ gives a straight line of critical values.
\end{proof}

\begin{figure}[p!]

\begin{minipage}[t]{\textwidth}

\begin{multicols}{2}

\begin{center}
\includegraphics[width=0.45\textwidth]{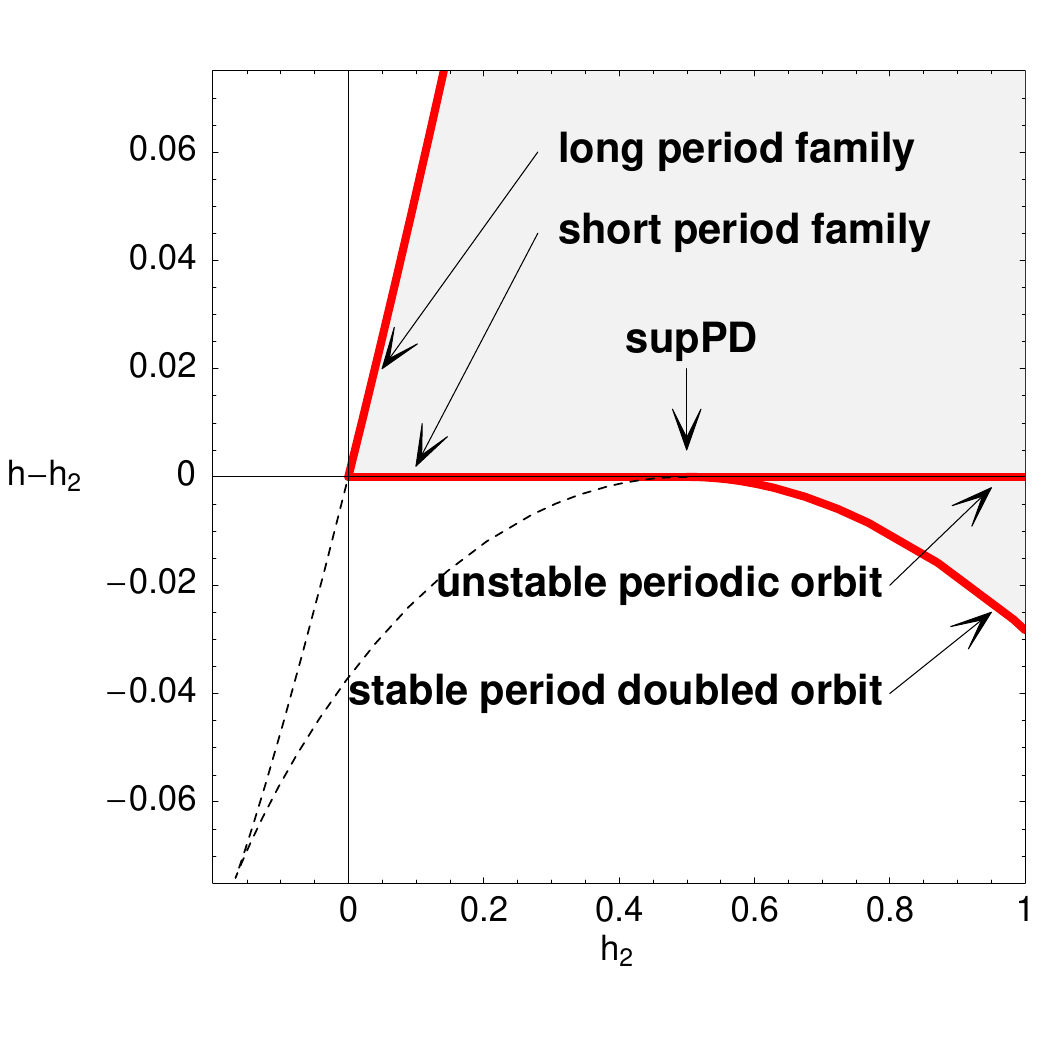}
\end{center}

\newpage

\begin{center}
\includegraphics[width=0.45\textwidth]{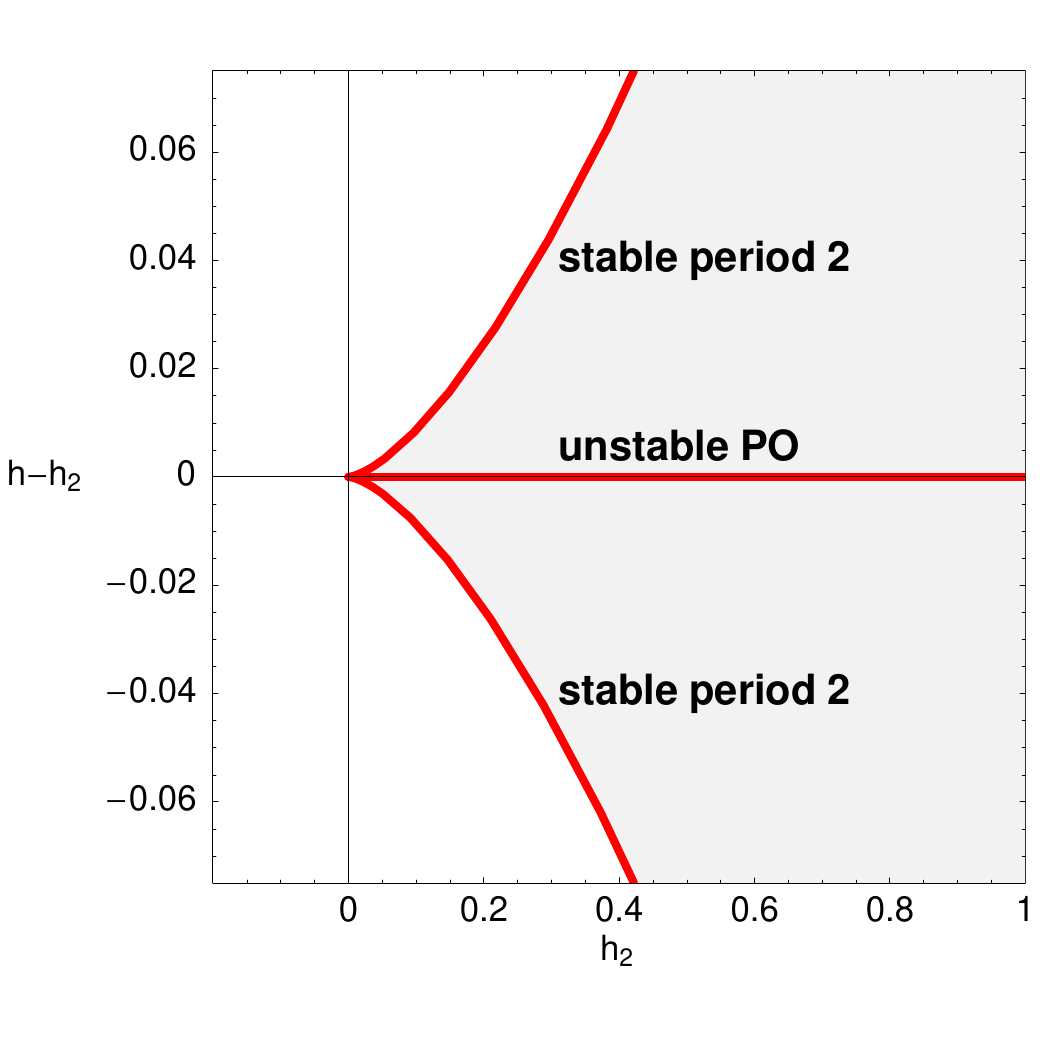}
\end{center}

\end{multicols}

\begin{multicols}{2}

\begin{center}
\includegraphics[width=0.45\textwidth]{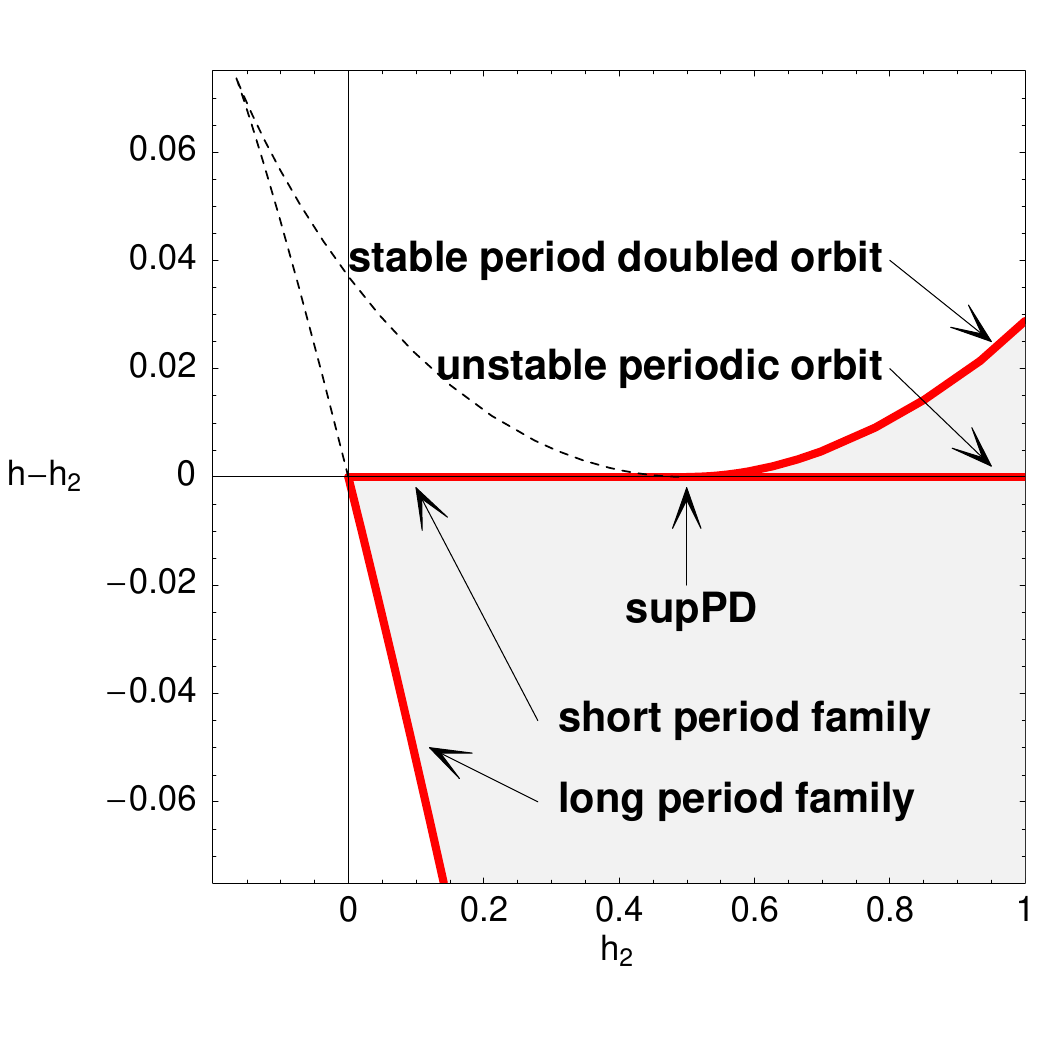}
\end{center}

\newpage

\begin{center}
\phantom{dummy}
\end{center}

\end{multicols}

\caption{\label{enermom:fig1}
{\small Image of the energy--momentum mapping for 1:2~resonance with $c=1$, $\lambda = 1/2$ (top left), $\lambda = 0$ (top right), $\lambda = -1/2$ (bottom).
Critical values are shown as thick red lines. Dotted lines are part of the discriminant locus,
but are outside the image of the energy--momentum mapping (shaded), which is only to the right
of the origin, bounded by the thick lines. The supercritical period doubling bifurcation is indicated by supPD.}
}
\end{minipage}

\end{figure}
% \newpage

\begin{figure}[p!]

\begin{minipage}[t]{\textwidth}

\begin{multicols}{2}

\begin{center}
\includegraphics[width=0.45\textwidth]{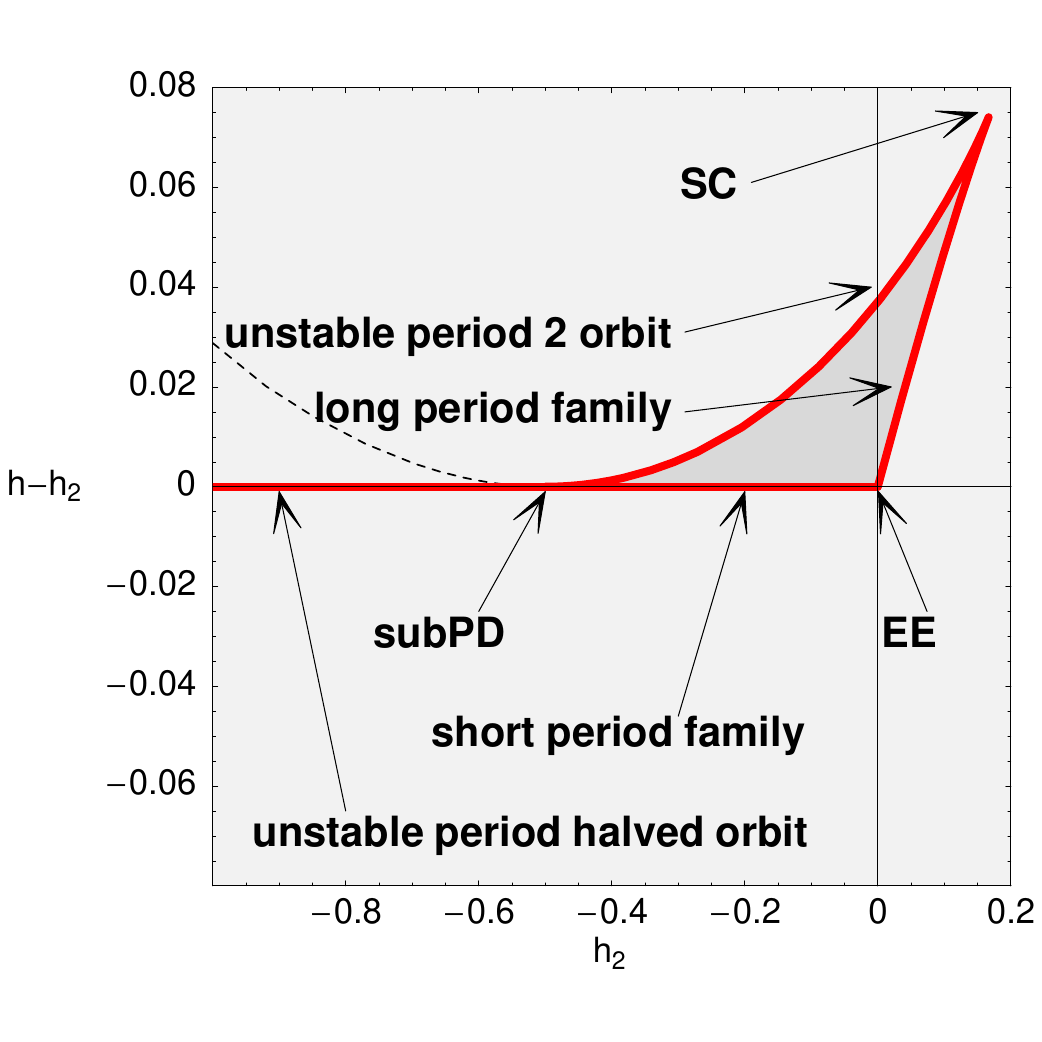}
\end{center}

\newpage

\begin{center}
\includegraphics[width=0.45\textwidth]{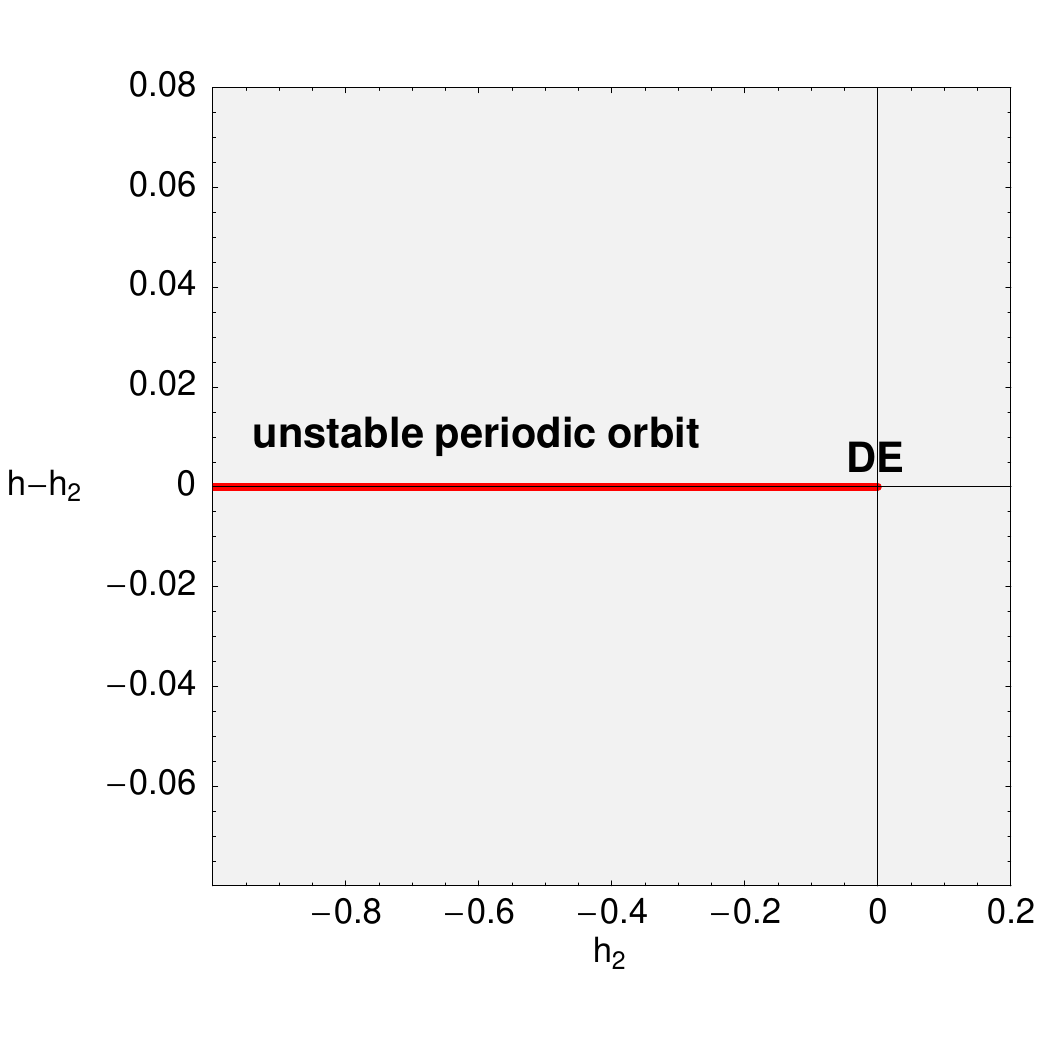}
\end{center}

\end{multicols}

\begin{multicols}{2}

\begin{center}
\includegraphics[width=0.45\textwidth]{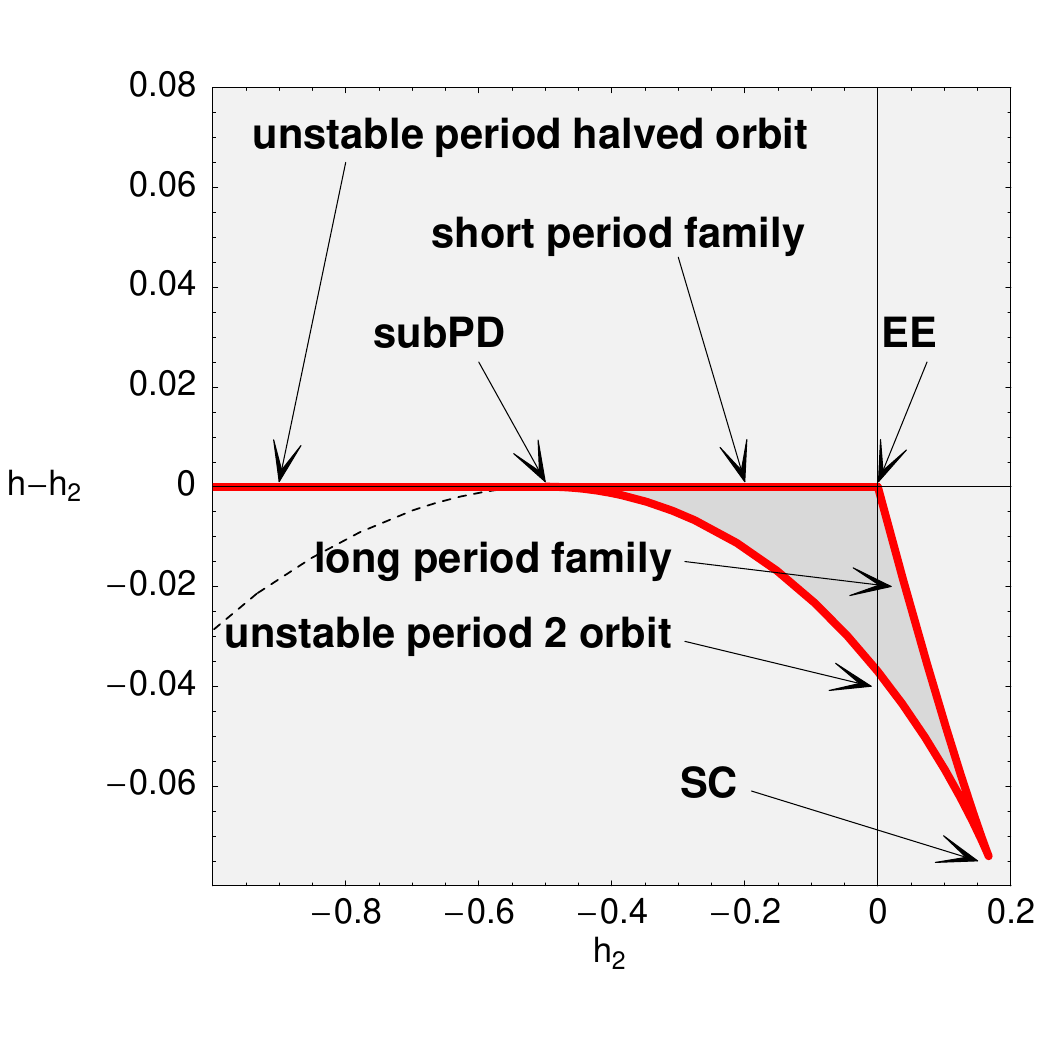}
\end{center}

\newpage

\begin{center}
\phantom{dummy}
\end{center}

\end{multicols}

\caption{\label{enermom:fig2}
{\small Image of the energy--momentum mapping for the $1$:$-2$~resonance with $c =1$, $\lambda = 1/2$ (top left), $\lambda = 0$ (top right), $\lambda = -1/2$ (bottom). Critical values are shown as thick red lines. Dotted lines are part of the discriminant locus, but not critical values of the energy--momentum mapping. The whole plane is the image of the energy--momentum mapping. The preimage in the dark shaded region due to detuning lifts to one two--torus $\T^2$ and a cylinder (figure \ref{reduction:fig3} middle), whereas it is one cylinder otherwise (figure \ref{reduction:fig3} top). Subcritical period doubling bifurcation, centre-saddle bifurcation, the elliptic equilibrium and the degenerate equilibrium are indicated by subPD, SC, EE and DE.}
}

\end{minipage}
\end{figure}

The characteristic feature of the curve of critical values of the energy--momentum mapping is that
%the curve
it always has a tangency with the straight line at $t = 0$.
However, the cusp is below the line $h = h_2$ for $\sigma \lambda > 0$,
and above for $\sigma \lambda < 0$. For $\lambda  = 0$ the cusp is on the line.

The transverse intersection at the origin marks the equilibrium point at the origin.
The tangency is a bifurcation of periodic orbits that are relative
equilibria of the reduced system. This bifurcation collides
with the equilibrium point at the passage through resonance when $\lambda = 0$.

Because $\pi_1 \geq 0$ the part of the discriminant locus given by $t \in [0, 2\lambda]$
is not part of the bifurcation diagram for $\sigma = 1$; similarly for $\sigma = -1$
it is the complement. Since $\pi_2 = \sigma (h_2 - \pi_1)/2 \geq 0$ and $\pi_1 \geq 0$ only the part of $h= h_2$ with $h_2 \geq 0$ is part of the bifurcation
diagram; similarly for $\sigma = -1$ it is the complement.

For $\sigma = 1$ the piece of the straight line between the transverse intersection
and the tangency is a stable relative equilibrium, beyond the tangency it is unstable.
The type of the bifurcation is determined by the multiplicity of the
preimage of the part of the curve beyond the tangency.
The solution for $\pi_i$ when $(h,g)$ are on the cubic curve is given by
\[
   \pi \;\; = \;\; \left(
   \frac{\sigma}{c^2} t(t - 2\lambda),
   \frac{(t - 2\lambda)^2}{4 c^2},
   \pi_1 a \frac{t-2\lambda}{2 c^2},
   \pi_1 b \frac{t-2\lambda}{2 c^2}
   \right) \;.
\]
A point in the pre--image is given by $p_1 = p_2 = q_2 = 0$ and
$\pi_1 = q_1^2/2$. The flow of $H_2$ acting on this point gives
all points in the pre--image. This is true for both signs of $\sigma$.
Hence this bifurcation always is a period doubling bifurcation.

The two branches emanating from the elliptic equilibrium (EE) at the
origin are always elliptic (they are the nonlinear normal modes).
For $\sigma = 1$ one of them stays stable; while the other one
loses stability in a supercritical period doubling bifurcation (supPD),
and creates another elliptic relative equilibrium with twice the period.
For $\sigma = -1$ one of them disappears in the centre--saddle
bifurcation (SC), while the other one collides with a
hyperbolic relative equilibrium in a subcritical period
doubling bifurcation (subPD) and only one hyperbolic relative equilibrium
(with half the period of the previous hyperbolic relative equilibrium)
remains. When $\lambda$ passes through zero, the r\^ole of the two
elliptic branches emanating from the equilibrium is exchanged.

Because $\sigma = -1$ gives a rotated critical curve the tangency now occurs for
negative $h_2$. The whole plane is in the image of the energy--momentum mapping
when $\sigma = -1$.
When $\lambda = 0$ the only critical values are those on the ray $h = h_2$
with $h_2 \leq 0$. This is the case studied in~\cite{ES04,CES} which has
fractional monodromy. Our unfolding shows that at the endpoint of the
ray ``half a swallowtail''  develops when the resonance is detuned.
By continuity, encircling this whole structure and crossing the
hyperbolic line $h = h_2$ for sufficiently large negative $h_2$
will give the same fractional monodromy.
In conclusion we have shown

\begin{theorem}
The unfolding of the 1:2~resonance shows that nearby there is a
supercritical period doubling bifurcation that passes through the
equilibrium at resonance.
The unfolding of the 1:$-2$~resonance shows that nearby there is a
subcritical period doubling bifurcation and a centre--saddle bifurcation
that pass through the equilibrium at resonance.
The unfolding of the 1:$-2$~resonant equilibrium point has fractional
monodromy.
\end{theorem}

\noindent
The proof of the last statement is deferred to Section~\ref{period}.

\subsection{Perturbation analysis}
\label{perturbation}

The equilibrium remains at the origin because the perturbation consists
of higher order terms. Persistence of the periodic orbits and their bifurcations is
straightforward. The Diophantine $2$--tori will persist, provided that the
Kolmogorov condition or iso--energetic non--degeneracy condition holds,
see the next sections. For $\sigma = -1$ we will show that the iso--energetic non--degeneracy
condition is violated along a line, while the Kolmogorov condition holds everywhere.
%Similarly
Similar to the focus-focus case \cite{HRDSanVuNgoc}, this can be considered
to be an effect of the monodromy of the equilibrium.

\section{Monodromy} % Period and rotation number
\label{period}

\begin{figure}[H]
\begin{center}
\includegraphics[width=0.55\textwidth]{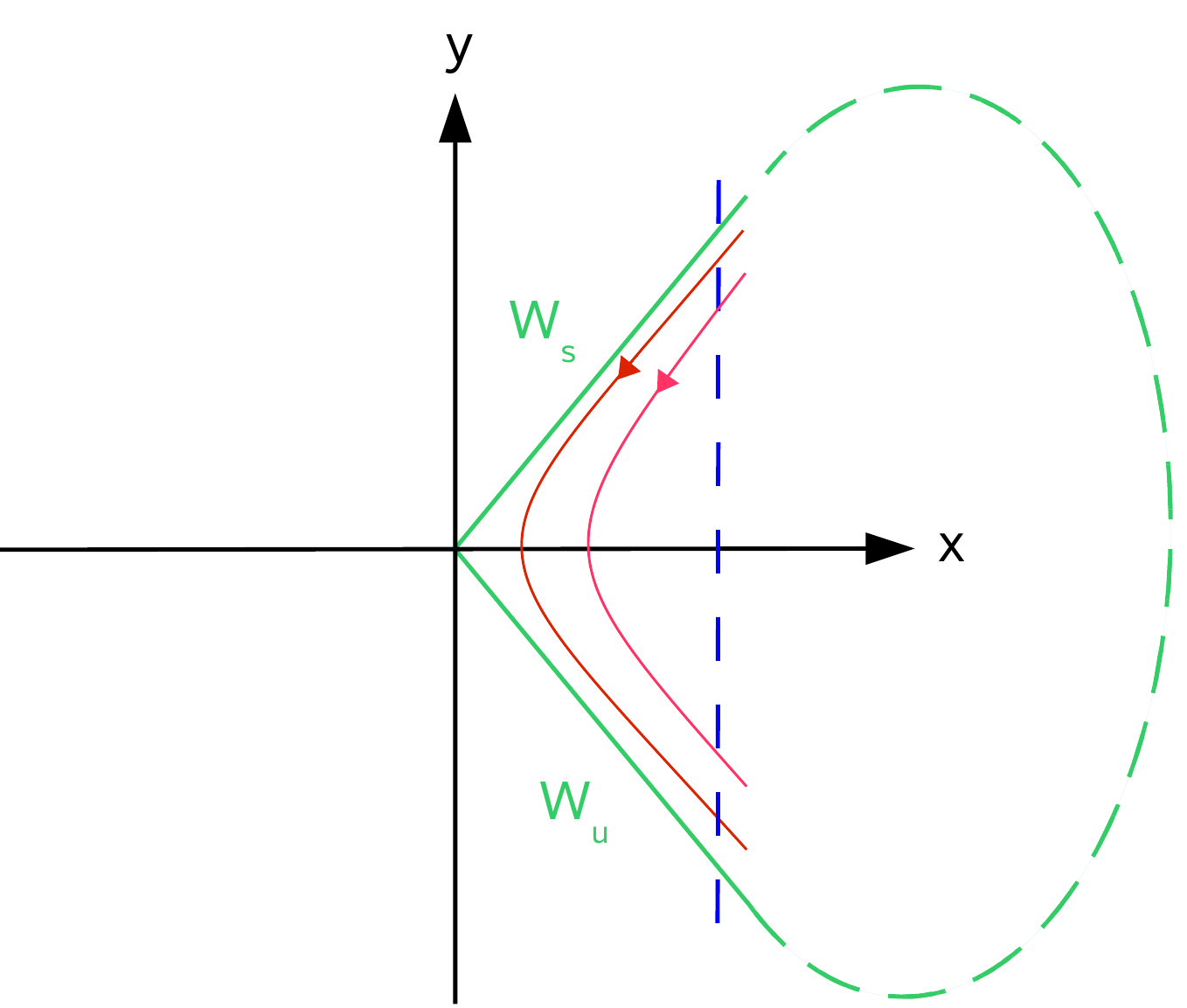}
\end{center}
\caption{\label{monodromy:fig1}
{\small Schematic representation of the dynamics for $\sigma=-1$ close to the separatrix of the singular point at the origin. The green dashed curve indicates the continuation of the stable and unstable manifold of the singular point when compactified.}
}
\end{figure}

The period of the reduced flow can be computed from
\[
    \dot \pi_1 = \{ \pi_1, \cH(\pi) \} = \{ \pi_1, \lambda \pi_1 + c \pi_3 \} = 2 c  \pi_4 \,.
\]
Fixing  $\cH = \deltah$ and using \eqref{Intro:Equ1},
$\pi^2_4$ becomes a polynomial in $\pi_1$.
Using ${\rm d}\pi_1 = w \, {\rm d}t$ the period integral is defined on the curve
\[
   \Gamma =  \left\{ (w, \pi_1) \mid
   w^2 =  Q(\pi_1) = 2 \sigma c^2 \pi_1^2 (h_2 - \pi_1) - 4 (\deltah - \lambda  \pi_1)^2 \right\} \,.
\]
It is given by
\[
   T(h_2, \deltah) \;\; = \;\;
   \oint_\gamma \frac{{\rm d}\pi_1}{w}  \;.
\]
For the compact case $\sigma = 1$ the meaning of this integral is clear.
We now focus on the case $\sigma = -1$.
A major problem in the non-compact case is to even define what is meant
by period, rotation number, and action.
Our approach is a simplified version of
%San V{\~u} Ng{\d o}cs~
\cite{SanVuNgocFF}. Assuming that there are no other critical points on the separatrix,
we can split the dynamics into two parts: singular dynamics near the critical point
and regular dynamics near the separatrix but away from the critical point, see figure
\ref{monodromy:fig1}.

Assuming that the system has compact invariant tori near the separatrix
the contribution of the dynamics far away from the equilibrium is some
bounded function with smooth dependence on the initial conditions
as long as we stay on the
same ``side" of the separatrix. This function contains the symplectic invariants of \cite{SanVuNgocFF}, but we are going to completely neglect it. As in \cite{SanVuNgocFF}, the closer we get to the critical value, the main contribution to the period comes from dynamics near the critical point. This contribution diverges in a characteristic way.
%and
It is this universal behaviour that we are interested in. For a hyperbolic point the period diverges logarithmically. Our equilibrium is not hyperbolic.
%, and we will instead
Instead we find
%algebraic divergence of
that the period diverges algebraically. There is one additional trick,
which has first been used in \cite{HRDIvanovSN}. Instead of integrating
along the orbit until it leaves a neighbourhood of the critical point, we integrate until the orbit reaches infinity. Thus the
%integrals become
integral involved becomes nice
%objects,
namely, a complete elliptic integral. The difference between integrating
up to some arbitrary cut-off point and
%between
integrating to infinity is
%again
a bounded smooth function, which can be neglected.
%and in particular bounded function in the present case, and thus
This becomes clear
%most clear
when considered from the point of view of algebraic geometry.
The frequency map is given by integrals
%over
of holomorphic one-forms over cycles on a complex affine ellpitic curve.
%closed curves.
After adding the point at infinity to make the affine elliptic curve a
compact elliptic curve in projective space, cycle that go to infinity on the affine curve are now compact. After this compactification the integral of the holomorphic
one form over the compact cycle gives the frequency map. The contribution near the point at infinity is smooth.
The asymptotics of the frequency map near the equilibrium point is obtained from the
contirbution of the cycle near the equilibrium point. This behaviour is unchanged
by adding higher order terms to the Hamiltonian.
%, and this is why
This is the reason why our analysis gives
the asymptotic behaviour for all completely integrable systems with the same normal
form up to order three.

The parameter $c$ can be removed by the scaling
$( \pi_1, h_2, \deltah ) \rightarrow 1/c^2( \pi_1, h_2, \deltah )$
(similar to the scaling in section 2).
The cycle $\gamma$ extends around the largest real root of~$w$ and infinity. This root has the same sign as $\deltah$. For values $(h_2, \deltah)$ near the line $h=h_2$ there are three real roots. The period integral is finite everywhere except when approaching the critical
% values on the non--positive $h_2$--axis, $T(h_2, 0) = \infty$ for $h_2 \leq 0$.
values along the line $h=h_2$, $T(h_2, 0) = \infty$ for $h_2 < 0$.
Even though the reduced motion is
%non--compact
unbounded, the time it takes to
reach~$\infty$ is finite. This
%is why now
explains why we can treat everything formally as if we
%would be
were in the compact case.

To compute the rotation number we can either differentiate the action $A$ (see below)
with respect to~$h_2$, or derive the differential equation for dynamics in the fibre.
We
%want to
will pursue the latter.

The $S^1$--action makes each cylinder a principal $S^1$--bundle, with
vertical direction along the fibre. Singling out a
%vertical
horizontal direction amounts to defining a connection.
While an angle in the fibre given by
\[
   \Theta \;\; = \;\;  \arg z_1
\]
is not well--defined, the $1$--form ${\rm d}\Theta$ does define a
principal connection simultaneously on all cylinders.
%The reason that we may not be able to define a zero--section and thus the angle
%conjugate to $H_2$,
%%itself
%is precisely that there is (fractional)
%monodromy --- the cylinder bundle is not trivial.
The formula $\{\Theta, H_2\} = 1$, however, can still can be used since
it only involves derivatives.
The Poisson brackets with $\Theta$, which describe the dynamics in the
fibre in
%full
phase space are computed to be
\[
\{ \Theta, \eta \} = 1, \quad
\{ \Theta, \pi_1\} = 1 , \quad
\{ \Theta, \pi_3\} = \pi_3/\pi_1
%\{ \Theta, \pi_4\} =
\,.
\]
Thus the dynamics of $\Theta$ generated by $\cH$ is given by
\[
  \Theta' \;\; = \;\; \{ \Theta, \cH \} \;\; = \;\;
     \{ \Theta, \lambda \pi_1 + c \pi_3 \}
   \;\; = \;\;  \lambda + \; c \, \frac{\pi_3}{\pi_1}
   = \lambda +  \frac{\deltah - \lambda \pi_1}{\pi_1}
   =   \frac{\deltah}{\pi_1} \,.
\]
For later use we define an integral that describes
%how much
the varation of $\Theta$
%changes according to the previous equation
when this dynamics in the base completes a full cycle. Changing the
integration from ${\rm d}t$ to ${\rm d}\pi_1$ and integrating
over a period $T$ gives the winding number as
\begin{equation}
   W(h_2, \deltah) = \frac{1}{2 \pi } \oint_\gamma \frac{\deltah}{\pi_1}
   						\frac{{\rm d}\pi_1}{w} .
\label{eq-windnum} \end{equation}
We call this integral the winding number, because
%it will turn out that
it measures the (fractional) monodromy of the system.\footnote{Even though a winding number is usually an integer,
%in this case
we will use the term for the function $W$ that changes by
a (half-) integer over a cycle around the origin of the energy-momentum map.
}
Note that this is not the rotation number of $\cH_3$, because
$\cH$ and $\cH_3$ differ by $H_2$.
This winding number integral (\ref{eq-windnum}) has a pole at $\pi_1 = 0$.
The residue of this pole is $\frac{1}{4 \pi} \I \, {\rm sgn} \, \deltah$.
% As a real integral over the allowed range of $\pi_1$  the integral needs to be multiplied by~$2$.
This integral shares many properties with the one discussed in \cite{CES}.
%The difference is that
However, our integral only describes the dynamics near the resonance,
while the integral in \cite{CES}
%also
the terms added for compactification also play a role. The singular behaviour near the equilibrium, however, is the same.

As already mentioned, for the computation of the rotation number of the
%this term is not define: true rotation number
full Hamiltonian \eqref{Reduction:Equ1} the term $H_2$ needs to be included.
The rotation number of the Hamiltonian without this term is what we defined
to be the winding number in the previous paragraph.
The equations of motion of the two Hamiltonians differ by $1$,
which originates from $\{ \Theta, H_2 \}= 1$.
Therefore the dynamics of $\Theta$ generated by the full Hamiltonian $\cH_3$ is given by
\[
   \dot \Theta = \{ \Theta, \cH_3 \}  =1 + \frac{\deltah}{\pi_1} \,.
\]
Changing the integration from ${\rm d}t$ to ${\rm d}\pi_1$ and integrating
over a period $T$ then gives the rotation number
\[
   2 \pi R(h_2, \deltah) = T(h_2, \deltah) +  2\pi W(h_2, \deltah) \;.
\]
%Now it is clear that the rotation number
%%will diverge
%diverges
%%as much
%as the period does.
%%why? The winding number integral may be unbounded too.
%
The period integral has logarithmic singularities. Therefore $R$ does as well.
For $W$ the situation is not so obvious.

In order to understand the behaviour of the integrals near the origin
we perform the scaling $\pi_1 = \sca^2 z$ and introduce scaled parameters
\[
      (h_2, \deltah, \lambda) \to (\sca^2 h_2, \sca^3 \deltah, \sca \lambda) \,.
\]
This is similar to the scaling done in section~2, but we repeat these steps here
because we want to interpret this scaling as choosing a new co--ordinate
system on the image of the energy-momentum map.
This scaling shows that the winding number is independent of the ``radius" $\sca$.
In order to study the behaviour of $W$ when the origin is encircled
we therefore introduce weighted polar co--ordinates in the image of the momentum map by
\[
      (h_2, \deltah, \lambda)  = (\sca^2 \cos\theta, \sca^3 \sin\theta, \sca \tilde \lambda)\,.
\]
The transformation from $(h_2, \deltah)$ to new variables near the origin of the momentum
map $(u,v) = \sca^{5/2}(\cos\theta, \sin\theta)$ is $C^0$ at the origin and $C^\infty$ everywhere else.
With this notation the crucial integrals become
\begin{equation} \label{TAWB}
  T(h_2, \deltah) = \frac{1}{\sca} A(\theta), \quad
  2 \pi W(h_2, \deltah) = B(\theta).
\end{equation}
If the degenerate equilibrium is approached on a curve
in parameter space $(h_2, \Delta h)$ with non-vanishing derivative at the origin, the leading order term of $T$ reads $\left| \Delta h \right|^{-1/3}$. Along the line
 $\Delta h=0, h_2>0$, we find $1/\sqrt{h_2}$ as leading order term of $T$. So in our two-degree of freedom system the leading order behaviour of the period $T$
as a function of $(h_2, \Delta h)$ depends on how the equilibrium is approached.
%which
This is in contrast to one degree of freedom systems where the energy $h$ is the only parameter.

\begin{figure}
\centerline{\includegraphics{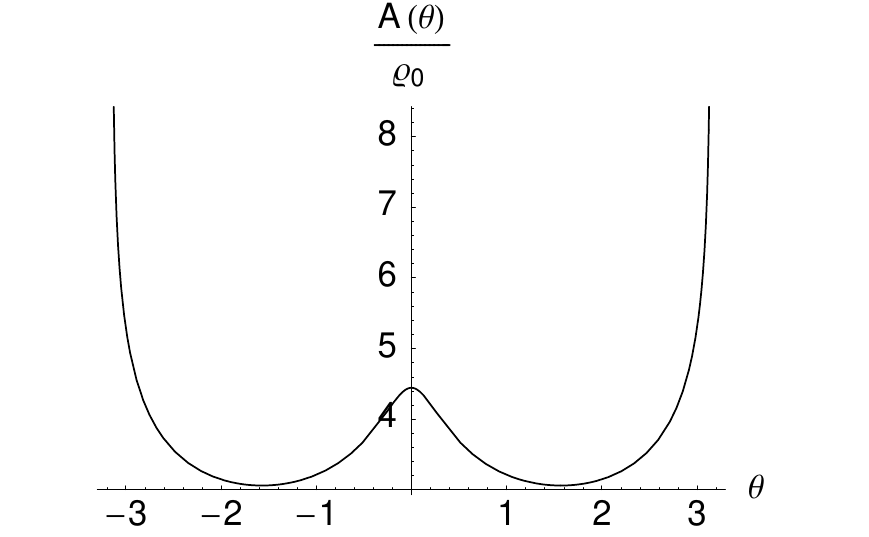}}
\centerline{\includegraphics{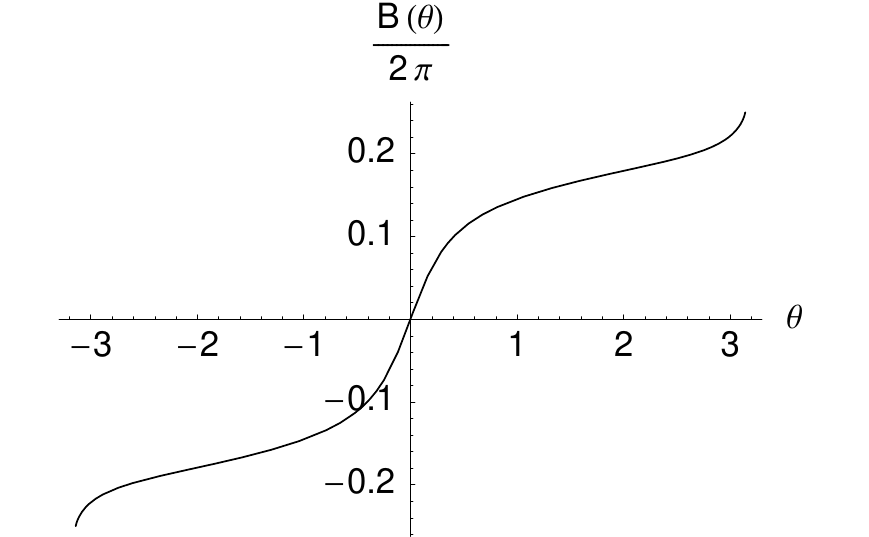}}
\caption{\label{monodromy:fig2}
{\small Period $T=A/\rho$ and winding number $W=B/2\pi$, see \eqref{TAWB}, as a function of $\theta \in [-\pi, \pi]$ for fixed $\sca = \sca_0$ and $\lambda = 0$. $T$ is even and periodic with logarithmic divergence at $\theta = \pm \pi$. $W$ is odd and continuous, but not periodic: it increases by $1/2$ when $\theta$ completes a circle.}
}
\end{figure}
%the axis labels $A$ and $B$ are not yet defined.

The integrals $A$ and $B$ are defined on the curve
\[
\tilde \Gamma = \{(z, w) :  \tilde w^2 = 2 \sigma z^2(\cos\theta - z) - 4(\sin\theta - \tilde \lambda z)^2 \}
\]
where
\[
  dA(\theta) = \frac{1}{\tilde w} d z, \quad
  dB(\theta) = \frac{\sin\theta }{z \tilde w} d z \,.
\]
The parameter $\tilde \lambda$ must be chosen sufficiently small so that
there are no critical values encountered on the loop around the origin in
the $(h_2, \deltah)$ plane. The plots of the functions $A$ and $B$ (i.e.\ of
$T$ and $W$ on a loop around the origin with $\sca = {\rm const}$)
are shown in figure~\ref{monodromy:fig2}.

The period $T$ diverges algebraically as $1 / \sca$ when the origin is approached.
This is due to the fact that the equilibrium is a degenerate
singularity: specifically, the value of the Hessian of the cubic integral
at the critical point (i.e. the origin) vanishes identially.
The winding number is independent of the distance from the origin $\sca$.
The period diverges logarithmically when $\theta \to \pm \pi$. This is the
usual divergence when approaching a hyperbolic periodic orbit.
The winding number is only a continuous function of $\theta $. It is
not differentiable. Surprisingly, the winding number
%The surprise is that it
is not a periodic function of $\theta$,
even though its defining integrand is.

We now prove that the increase in values of $W$ (or $B$)
when we let $\theta \in [-\pi, \pi]$ is indeed as sugested by
figure~\ref{monodromy:fig2}.
First of all observe that $B$ is an odd function,
so we can restrict to $\theta \ge 0$. For $\lambda = 0$ the polynomial $\tilde w^2$
has one non-negative real root and two roots with non-positive real part (which may
be real or complex). For $\theta = 0$ these are $(1,0,0)$ and for $\theta = \pi$ these
are $(0,0,-1)$. There is a double root for $\theta  =0$. But since we always
integrate from the largest root onwards this does not concern us
(there is another collision of roots, but again it is not in the physically allowed region). Thus the function $B(\theta)$ is smooth for $\theta \in (-\pi, \pi)$.
The double root at 0 for $\theta =\pi$ is at the boundary of the integration interval,
so care needs to be taken when evaluating $B$. At first sight it looks like
$B(\pi) = 0$ (which would also be required for a periodic odd function).
However, we will now show that $B(\theta)$ is not periodic, but instead
has non-zero $\pi/2$ value at $\theta = \pi$. Since $B$ is odd therefore
$B(-\pi) = -\pi/2$ and $B(\pi) - B(-\pi) = \pi$: the value of $B$ increases
by $\pi$ when $\theta$ traverses $[-\pi, \pi]$.

\begin{Lemma} \label{quarter}
\[
	\lim_{\theta \to \pm \pi^\mp } B(\theta) =  \pm \pi/2 + \arcsin \left( \sqrt{2} \mspace{2mu} \lambda \right)
\]
\end{Lemma}
\begin{proof}

We consider the limit to $+\pi$ from the left first.
When $\theta \to \pi$ we can treat $\delta = \sin \theta$ as a small parameter.
The colliding roots behave like $\delta$, thus we scale $z = y \delta$ and find
\[
dB = 1 / (y \sqrt{-2y^2(\cos\theta - y \sin\theta ) - 4(1-\lambda y)^2}) dy
\]
The root that limits to $z = -1$ in the new variables tends to $-\infty$ when $\delta \to 0$. The other two roots of opposite sign tend to non-zero values.
Denote these roots (for any small value of $\delta$) by $y_\infty<y_- < y_+$.
The integration interval now is $[y_+, \infty)$. In order to regularize the square root singularity of the integrand at $y_+$ and also make the lower integration boundary independent of $\delta$ we introduce new variables by $y = y_+ + u^2$. Now the integrand is a smooth function of $u$ and $\delta$ uniformly over the integration interval. We now can compute the Taylor series
in $\theta - \pi$. The first term gives the value of $B$ at $\theta = \pi$ as
%\[   B = \frac{1}{2\pi} 2 \int_0^\infty \frac{\sqrt{2} \, {\rm d} u}{(\sqrt{2} + u^2)\sqrt{2\sqrt{2} + u^2}} = \frac14\]
\[
    B(\pi) = 2 \int_0^\infty \frac{\sqrt{2} \, {\rm d} u}{(u^2 + y_+)\sqrt{2y_+ + 4\lambda(1-\lambda y_+) + u^2(1-2\lambda^2)}}
\]
Since the integrand is even function of $u$, the interval of
integration can be extended to $[-\infty, \infty]$, absorbing one factor of 2. Direct evaluation of the integral gives the stated result. The computation is similar for $B(− \pi)$, except that now $\delta$ is negative which changes the sign of the constant term in the result, but not the $\lambda$-dependent part. As a result the difference $B(\pi) - B(-\pi) = \pi$ is independent of $\lambda$.
\end{proof}

For later use we also record the following

\begin{lemma} \label{monotone}
The function $B(\theta)$ for $\lambda = 0$ is strictly
%monotonous.
monotonic.
\end{lemma}

\noindent \textit{Discussion}
In the appendix we give the expression for $B'$ in terms of Legendres
complete standard integrals $K$ and $E$. Using these formulas it can
be numerically unambiguously verified
%maybe some hint at the numerical precision used should be given. Since
%exponentially small phenomenon can occur 100 digits of accuracy may be needed.
that $B'$ does not have a zero.
The function $B'$ can be plotted with any standard symbolic algebra program.
We find $B' > 0.15$, so it is bounded away from zero such that only a few
digits of accuracy are needed.
Thus round-off errors in the computation do not play a role.
We have not been able to find an analytic proof  yet.
\hfill \textit{end of discussion}

To establish monodromy is now simply a matter of finding an expression for
the actions $I = (I_1, I_2)$ for the integrable system. One action
$I_2$ is trivial, namely
the generator of the $S^1$ symmetry. Set $I_2 = h_2$.
The non-trivial action $I_1$ of the system (or more precisely, its leading order term
near the equilibrium) contains some terms that depend
smoothly and periodically on $\theta$.
%but then
However, it has a term $W h_2$,
%in which the trivial action $h_2 = I_2$ is multiplied by the winding number,
which changes by $\frac{1}{2}$ over a cycle
around the origin,
%from
see Lemma~\ref{quarter}. Thus $I_1$
%will be changed by
changes by $\frac{1}{2} \, I_2$
%upon
after traversing this loop. This behaviour
%has been
is called fractional monodromy \cite{NSZ}.

\begin{lemma} \label{action}
The non-trivial action $I_1$ of the 1:$-2$ resonance has the form
\[
     I_1(h_2, \deltah) =  \frac32 \frac{\deltah}{2 \pi} T - h_2 W \,.
\]
It is discontinuous across the singular value at $\deltah = 0$ and it changes
by $h_2 / 2 = I_2 /2$ along a loop around the origin.
\end{lemma}

\begin{proof}
For the computation of the action of the reduced system two canonically
conjugate functions whose Poisson bracket equals one are needed.
It is easy to check that
\[
   \{ \cos^{-1} \frac{\pi_3}{\sqrt{\pi_1^2 \pi_2 }}, \pi_2 \} = 1
\]
before introducing $\eta$ in the Poisson structure. Now the action is given by
\[
I_1(h_2, \deltah) = -\frac{1}{2\pi} \oint \cos^{-1} \left( \frac{\pi_3}{\sqrt{\pi_1^2 \pi_2 }}
 \right) \, \dee \pi_2 .
\]
Integration by parts and changing the integration variable to $\pi_1 = h_2 + 2 \pi_2$ gives
\[
I_1(h_2, \deltah) = \frac{\deltah}{2\pi} \oint  \left( \frac32 - \frac{h_2}{\pi_1} \right) \frac{\dee \pi_1}{w}
\]
and the result follows.
\end{proof}
The formula of Lemma~\ref{action} should be compared to the analogous
expression for a focus-focus point, which is $I_1 = J_1 T - J_2 W$ where
$J_i$ are the co--ordinates from the Eliasson normal form \cite{SanVuNgocFF}. The two formulas are surprisingly similar, even though the corresponding equilibria are quite different.

Note that the remarkable formula of Lemma~\ref{action} can also be read
as a decomposition of the angle change $2\pi W$ into a geometric phase (namely
the action) and a dynamic phase (the term proportional to $T$).
In many cases simple relations like this can be found for the change of angle
in the fibre computed from quantities in the base, see, e.g., Montgomery~\cite{Montgomery90}.

For us the most important aspect of Lemma~\ref{action} is that together with
Lemma~\ref{quarter} it immediately implies
the following monodromy theorem,
complementing the work of \cite{NSZ,CES}.

\begin{theorem}
The Hamiltonian system describing the 1:$-2$ resonance has fractional monodromy
$\frac{1}{2}$.
\end{theorem}

Let us point out the philosophy in our approach once more.
%On the one hand following Duistermaat \cite{Duistermaat1980} classical monodromy can be %understood by considering the period lattice, or equivalently, following Cushman \cite{} the %rotation number of the system. On the other hand, following Duistermaat and Cushman \cite{},
%quantum monodromy is associated with the behaviour of the classical actions.
Following Duistermaat \cite{Duistermaat1980} classical monodromy can be understood by considering the period lattice, or equivalently, following Cushman \cite{CushmanBates1997} the rotation number of the system. Following Zhilinski\'i \cite{Zhilinskii2006} another way to think of Hamiltonian monodromy is in terms of lattice defects in the associated quantum analog. In the present case the rotation number is singular at the critical values of the energy-momentum map along the line $h=h_2$, and the fibres are unbounded. This makes it hard to speak about continuing the period lattice or rotation number
around the origin (the degenerate critical point corresponding to the equilibrium).
The first papers discovering fractional monodromy \cite{NSZ} used a geometric approach.
Later \cite{ES04} a rescaled time was used in order to ``remove" the blowup of $T$.
Our approach is similar to this, in that we are dealing with a similar type
of elliptic integrals. There are two differences. Firstly, in the proof of monodromy we directly consider the actions of the system. Even when the rotation number is singular, the action can always be chosen continuous across separatrices (up to integer factors counting the number of tori, etc.). In our case the action can be made continuous at any point, but globally
it is not single valued, due to the contribution of $W$.
It seems that in general the actions are the more ``robust" objects that are
well suited for this type of consideration.
Secondly, instead of compactifying the Hamiltonian by adding higher
order terms we directly treat the non-compact situation, by only considering
dynamics near the singularity.
This is inspired by the approach of San Vu Ngoc \cite{SanVuNgocFF} to the symplectic
invariants near focus-focus points.
Our main new results are presented in the next section, where we consider the
derivatives of the functions $A$ and $B$ in order to study the
non-degeneracy condition of the KAM theorem.

\section{Frequency Map}
\label{freqMap}

The results in this and the following section are valid at the bifurcation
$\lambda = 0$ only.
For perturbation theory we have to consider the actual rotation number $R$
not just the less singular winding number part.
To study the derivatives of $R$ and the frequencies it is best to use the
scaled forms of the integrals. Only the leading term when $\sca \to  0$ is
relevant. Inverting $\partial(h_2, h)/ \partial (\sca, \theta)$ gives
\[
   \frac{\partial \sca}{\partial h_2} = -\Delta  \sca^2 \sin\theta + O( 1/\sca ) , \quad
   \frac{\partial \theta}{\partial h_2} = -2 \Delta \sca  \cos\theta + O( 1/\sca^2 )\,,
\]
where the determinant of the transformation is
\[
   \Delta = \left| \frac{ \partial (\sca, \theta) } { \partial(h_2, h)}\right| =
   \frac{2}{\sca^4(5- \cos2\theta)} \,.
\]

\begin{theorem} (Vanishing Twist)
The isoenergetic non-degeneracy condition is violated on a line emerging from
the critical value at the origin of the 1:$-2$ resonance. Namely
the function $R(h_2)$ has a critical value at $h_2=h$ for $\lambda = 0$.
\end{theorem}

\begin{figure}
\centerline{\includegraphics[width=0.5\textwidth]{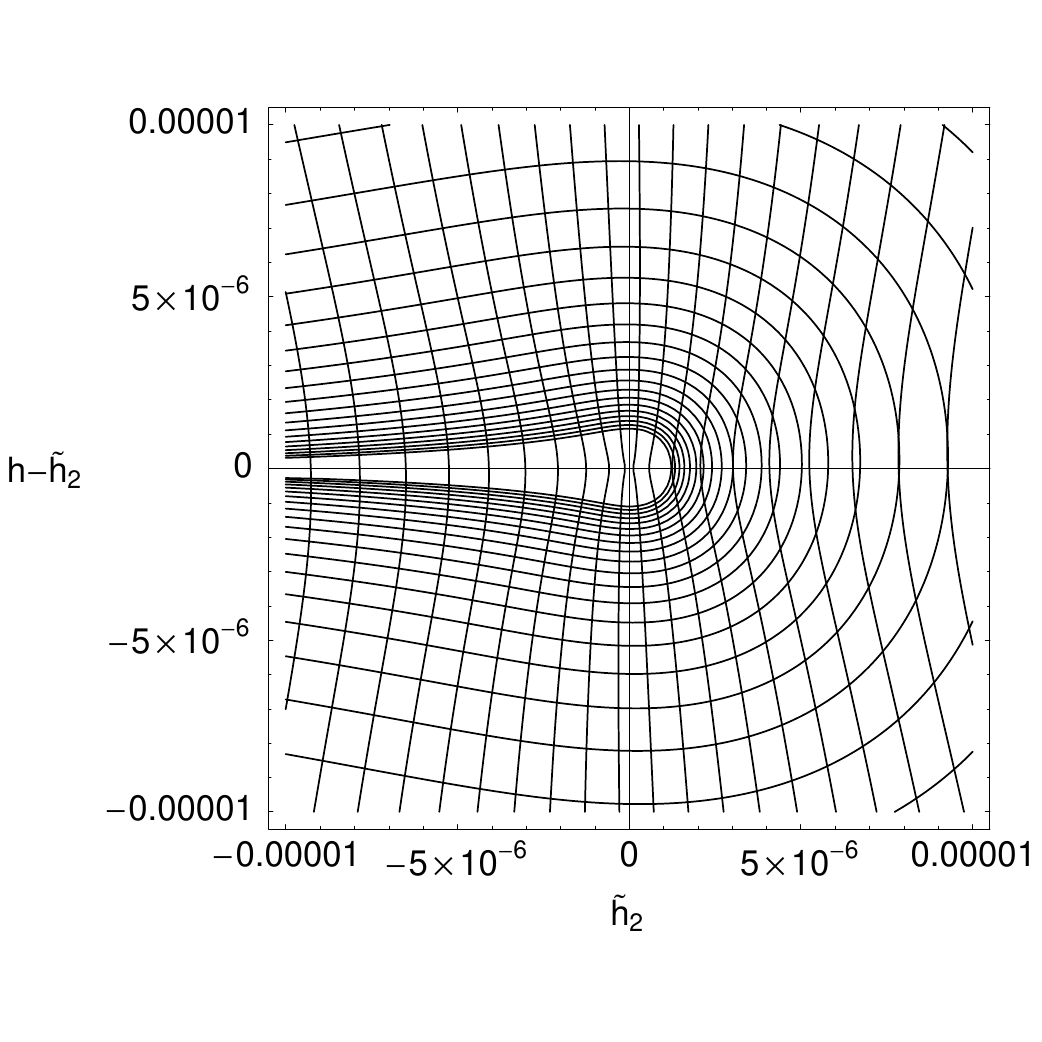}}
\caption{\label{freqMap:fig1}
{\small Illustration of the condition on vanishing twist in the $1:-2$ resonance. The twist condition is violated where the lines of constant $h$ are tangent to the contour lines of $R(h_2,\deltah)$ which is the case for $\deltah = 0$ and $\tilde{h}_2 > 0$. $\tilde{h}_2 = \sca^3 \cos \theta$.}
}
\end{figure}

\begin{proof}
Computing the derivative of $2 \pi R = A(\theta)/\sca + B(\theta)$ after scaling gives
\[
2 \pi \frac{\partial R}{\partial h_2} = -\frac{A}{\sca^2} \frac{\partial \sca}{\partial h_2}
+ \left(\frac{A'}{\sca} + B' \right) \frac{\partial \theta}{\partial h_2}
= \Delta (A \sin\theta - 2 A' \cos\theta ) + O(1/\sca^3)
\]
For $\lambda = 0$ the function $A$ is even in $\theta$ and therefore the leading order term
vanishes at $\theta = 0$. For $\lambda \not = 0$ the origin becomes a non-degenerate
elliptic fixed point, and the vanishing twist will instead be present near the centre--saddle bifurcation \cite{HRDIvanovSN}, see the bifurcation diagram figure~\ref{enermom:fig2}.
\end{proof}
A graphical illustration of the condition on vanishing twist is given in figure \ref{freqMap:fig1}.

% \subsection{Kolmogorov non-degeneracy}

The frequencies of the system are
\[
   \omega_1 = 2\pi/T, \quad \text{and} \quad \omega_2 = 2\pi R/T= 1 + 2 \pi W/T \,.
\]
%Denote the actions of the system by $I = (I_1, I_2)$. Here $I_2 = h_2$,
%while $I_1$ is given by some elliptic integral.

\begin{theorem}
The Kolmogorov non-degeneracy condition
\[
     \left| \frac{\partial \omega}{\partial I} \right| \not = 0
\]
is satisfied at {\em every} point in a neighbourhood of the origin of the 1:$-2$ resonance.
\end{theorem}

\begin{proof}
The determinant is the product of three determinants
\[
   \left| \frac{\partial \omega}{\partial I} \right| =
      \left| \frac{\partial \omega}{\partial (\sca, \theta) } \right|
      \left| \frac{\partial (\sca, \theta)} {\partial (h_2, h) }\right|
      \left| \frac{\partial (h_2, h)}{ \partial (I_1, I_2)} \right|
\]
The middle one is $\Delta$, and hence non-zero outside the origin.
Since $h_2 = I_2$ the last one reduces to $- \partial h/\partial I_1 = - \omega_1$
which is non-zero outside the critical values.
The non-trivial first determinant is (exactly!) given by
\[
      \left| \frac{\partial \omega}{\partial (\sca, \theta) } \right|
=
      \left| \begin{pmatrix} 2\pi/A & -2\pi\sca A'/A^2 \\ B/A & \sca(B'A - A'B)/A^2 \end{pmatrix} \right|
= 2 \pi \sca \frac{B'}{A^2} \,.
\]
In  Lemma~\ref{monotone} we have shown that $B(\theta)$ is
%monotonous
monotonic, see figure~\ref{monodromy:fig2}. This completes the proof. \medskip

%Altogether
Collecting the above results we find that
\[
     \left| \frac{\partial \omega}{\partial I} \right|
     =  - (2 \pi \sca)^2 \frac{B'}{A^3}  \Delta
\]
Notice that this diverges like $1/\sca^2$ when approaching the origin
away from $\theta = \pm \pi$. When $\theta \to \pm \pi$ for fixed $\rho$, however,
the determinant approaches zero because both, $A$ and $B'$
diverge logarithmically.
\end{proof}

\begin{figure}
\centerline {\includegraphics[width=0.45\textwidth]{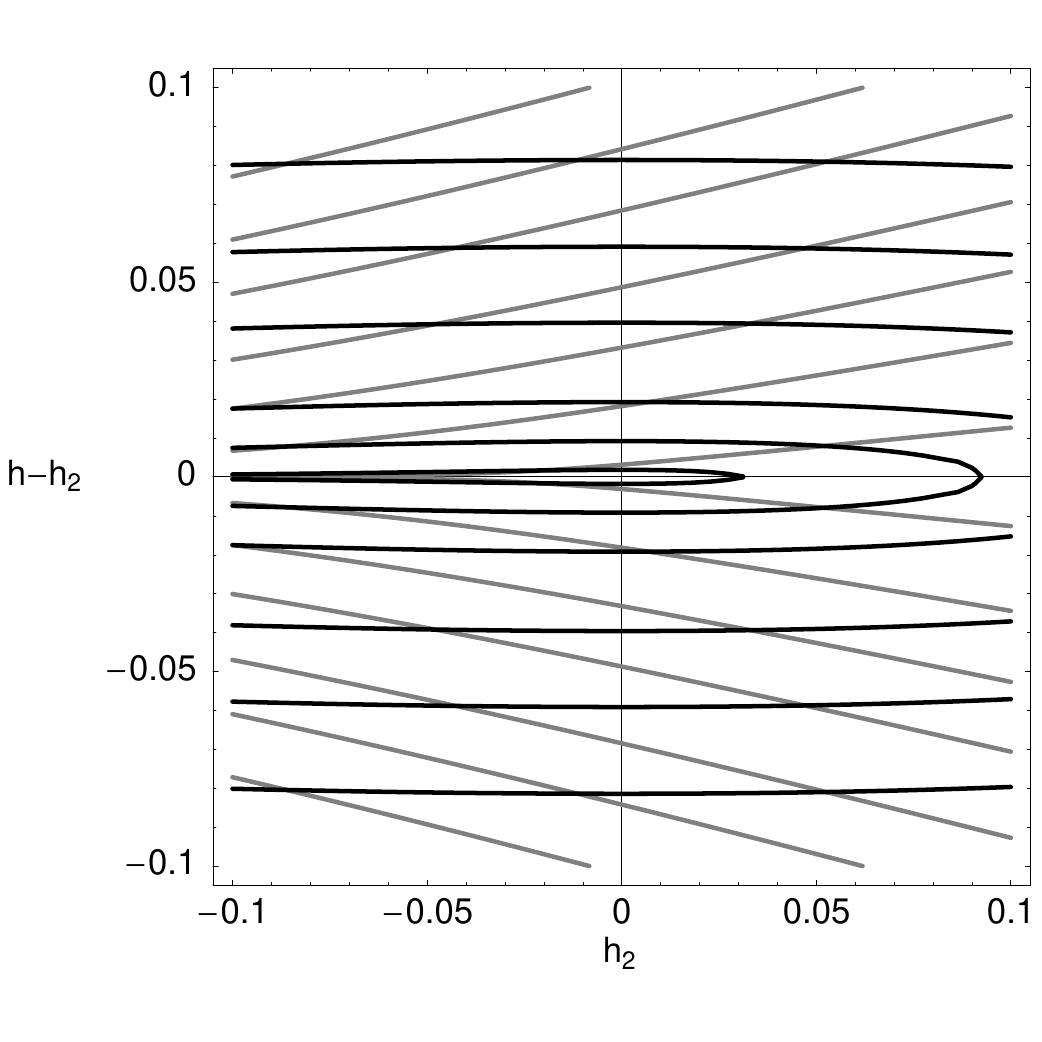}}
\caption{\label{freqMap:fig2}
{\small The non--violation of the Kolmogorov condition means the frequencies
plotted in the image of the energy--momentum mapping give a ``nice'' co--ordinate system. The contour lines of $\omega_2$ are black, the ones of $\omega_1$ gray.}
}
\end{figure}

In this section we only considered the case without detuning, $\lambda = 0$.
However, using the analysis for period doubling in \cite{DullinIvanov2004} and for the centre--saddle bifurcation in \cite{HRDIvanovSN} it is straightforward to
get a qualitative understanding of vanishing twist for $\lambda \not = 0$ as well.
For $\sigma = +1$ no vanishing twist is forced by the bifurcation.
For $\sigma = -1$ the cusp of the centre--saddle bifurcation will
produce a curve emanating from it along which the twist condition is violated.
For $\lambda \to 0$ this curve will locally coincide with $\deltah = 0$ as
predicted here. Combining the images of the energy-momentum map for
different parameters $\lambda$ into a three-dimensional picture there
will therefore be a surface along which the twist condition is violated.
To our knowledge the Kolmogorov condition has not been investigated near
these bifurcations.

\subsection*{Acknowledgements}
HH thanks Hans Duistermaat for helpful discussions.
RHC, HRD and HH thank the Centre Interfacultaire Bernoulli at the
EPF Lausanne for its hospitality.
This research was partially supported by the European Research Training Network
{\it Mechanics and Symmetry in Europe\/} (MASIE), HPRN-CT-2000-00113.
HRD acknowledges partial support by a research fellowship of the Leverhulme trust.

\appendix

\section{Standard Form of the Integrals $A$, $B$, $A'$, and $B'$}

Here we present the main elliptic integrals in Legendre normal form for the cases of three real and one real root of $\tilde w^2$. This can be avoided using a Landen transformation \cite{PFByrdMDFriedman1971} but since this doesn't simplify the Legendre normal forms we refrained from doing so. We denote the real roots by $a > b > c$ and the complex conjugate ones by $b$ and $\bar{b}$.

$A(\theta)$ is simply given by
\begin{eqnarray}
	A(\theta) = \sqrt{2} \, g \, K(m)
\end{eqnarray}
where
\begin{eqnarray*}
g = \frac{2}{\sqrt{a-c}}, \quad m = \frac{b-c}{a-c}
\end{eqnarray*}
for three real roots and
\begin{eqnarray*}
g = \frac{1}{\sqrt{d}}, \quad m = \frac{d + b_1 - a}{2 \, d}, \quad
d^2 = ( \re b - a )^2 + (\im b)^2
\end{eqnarray*}
for one real root.
The derivative of $A$ is of 2nd kind. For three real roots we find
\begin{eqnarray}
	A'(\theta) = \sqrt{2} g \left[ K(m) \left( u_1 + a \, u_2 \right) - u_2 (a-c) E(m) \right]
\end{eqnarray}
with
\begin{alignat*}{3}
u_1 &= \frac{\sin 2 \theta}{2 D} \left( \mu^2 \cos \theta - 9 \right), \quad &
u_2 &= \frac{\mu^2 \cot \theta}{4 D} \left( \cos 2 \theta - 5 \right)\,,
\end{alignat*}
where $D$ is the discriminant of $\tilde{\omega}^2$ given by
$D = 2 \mu^2 \cos^3 \theta + 27 \sin^2 \theta$.
In the case of one real root we have
\begin{eqnarray}
	A'(\theta) = \sqrt{2} g \left[ K(m) \left( u_1 + ( a + d ) \, u_2 \right)
	- 2 \, d \, u_2 \, E(m) \right].
\end{eqnarray}

$B(\theta)$ is the elliptic integral of 3rd kind
\begin{eqnarray}
	B(\theta) = \sqrt{2} \, g \, \sin \theta \left[ \frac{K(m)}{b}
	+ \frac{b-a}{b \, a} \Pi \left( \frac{b}{a}, m \right) \right]
\end{eqnarray}
in the case of three real roots and
\begin{eqnarray}
	B(\theta) = \sqrt{2} \frac{g}{a + d} \, \sin \theta \left[ \frac{K(m)}{\alpha}
	+ \frac{\alpha - 1}{\alpha (1-\alpha^2)} \Pi \left( \frac{\alpha^2}{\alpha^2 - 1}, m \right) \right]
\end{eqnarray}
otherwise, where
\begin{eqnarray*}
	\alpha = \frac{a - d}{a + d}.
\end{eqnarray*}
Its derivative is of 2nd kind. For three real roots we find
\begin{eqnarray}
	B'(\theta) = - \sqrt{2} \, g \, \left[ K(m) \left( u_1 + a \, u_2 \right)
	- u_2 (a-c) E(m)   \right]
\end{eqnarray}
where
\begin{alignat*}{3}
u_1 &= \frac{\mu^2 \cos \theta}{4 D} \left( \cos 2 \theta -  5 \right), \quad &
u_2 &= \frac{3 \mu^2}{4 D} \left( 2 + \sin^2 \theta \right).
\end{alignat*}
In the case of one real root we finally have
\begin{eqnarray}
	B'(\theta) = - \sqrt{2} \, g \, \left[ K(m) \left( u_1 + ( a + d ) \, u_2 \right) - 2 \, d \, u_2 \, E(m) \right].
\end{eqnarray}

\bibliographystyle{plain}
\bibliography{newtex1to2}

 \end{document}